\documentclass{emulateapj}

\usepackage{amsmath}
\usepackage{amssymb}
\usepackage{graphicx}
\usepackage{color}
\usepackage{natbib}
\usepackage{epsfig}

\def\gtaprx {\lower .1ex\hbox{\rlap{\raise .6ex\hbox{\hskip .3ex
	{\ifmmode{\scriptscriptstyle >}\else
		{$\scriptscriptstyle >$}\fi}}}
	\kern -.4ex{\ifmmode{\scriptscriptstyle \sim}\else
		{$\scriptscriptstyle\sim$}\fi}}}
\def\ltaprx {\lower .1ex\hbox{\rlap{\raise .6ex\hbox{\hskip .3ex
	{\ifmmode{\scriptscriptstyle <}\else
		{$\scriptscriptstyle <$}\fi}}}
	\kern -.4ex{\ifmmode{\scriptscriptstyle \sim}\else
		{$\scriptscriptstyle\sim$}\fi}}}

\newcommand{\cutt}[1]{\textcolor{blue}{}}

\newcommand{\Ms}{${M}_{\odot}$}

\newcommand{\Ni}{{\ensuremath{^{56}\mathrm{Ni}}}}

\begin{document}

\title{Supermassive Population III Supernovae and the Birth of the First Quasars}

\author{Daniel J. Whalen\altaffilmark{1,2,3}, Wesley Even\altaffilmark{4}, Joseph 
Smidt\altaffilmark{1}, Alexander Heger\altaffilmark{5}, K.-J. Chen\altaffilmark{6}, Chris 
L. Fryer\altaffilmark{4},  Massimo Stiavelli\altaffilmark{7}, Hao Xu\altaffilmark{8} and 
Candace C. Joggerst\altaffilmark{9}}

\altaffiltext{1}{T-2, Los Alamos National Laboratory, Los Alamos, NM 87545}

\altaffiltext{2}{McWilliams Fellow, Department of Physics, Carnegie Mellon University, 
Pittsburgh, PA 15213}

\altaffiltext{3}{Universit\"{a}t Heidelberg, Zentrum f\"{u}r Astronomie, Institut f\"{u}r 
Theoretische Astrophysik, Albert-Ueberle-Str. 2, 69120 Heidelberg, Germany}

\altaffiltext{4}{CCS-2, Los Alamos National Laboratory, Los Alamos, NM 87545}

\altaffiltext{5}{Monash Centre for Astrophysics, Monash University, Victoria, 3800, Australia}

\altaffiltext{6}{School of Physics and Astronomy, University of Minnesota, Twin Cities, 
Minneapolis, MN  55455}

\altaffiltext{7}{Space Telescope Science Institute, 3700 San Martin Drive, Baltimore, MD 21218}

\altaffiltext{8}{Center for Astrophysics and Space Sciences, UC San Diego, La Jolla, CA  92093}

\altaffiltext{9}{XTD-3, Los Alamos National Laboratory, Los Alamos, NM 87545}

\begin{abstract}

The existence of supermassive black holes (SMBHs) as early as $z \sim$ 7 is one of the great, 
unsolved problems in cosmological structure formation.  One leading theory argues that they 
are born during catastrophic baryon collapse in $z \sim$ 15 protogalaxies that form in strong 
Lyman-Werner (LW) UV backgrounds.  Atomic line cooling in such galaxies fragments baryons 
into massive clumps that are thought to directly collapse to 10$^4$ - 10$^5$ \Ms\ black holes.  
We have now discovered that some of these fragments can instead become supermassive 
stars that eventually explode as thermonuclear supernovae (SNe) with energies of $\sim$ 10$
^{55}$ erg, the most energetic explosions in the universe.  We have calculated light curves and 
spectra for supermassive Pop III SNe with the Los Alamos RAGE and SPECTRUM codes.  We 
find that they will be visible in near infrared (NIR) all-sky surveys by \textit{Euclid} out to $z \sim$ 
10 - 15 and by \textit{WFIRST} and \textit{WISH} out to $z \sim$ 15 - 20, perhaps revealing the 
birthplaces of the first quasars.

\end{abstract}

\keywords{black hole physics - cosmology: early universe - theory - galaxies: formation -- galaxies: 
high-redshift -- stars: early-type -- supernovae: general -- radiative transfer -- hydrodynamics -- 
shocks}

\maketitle

\section{Introduction}

One model for the origin of SMBHs, which have now been found at $z \sim$ 7, or less than 
a Gyr after the big bang \citep{mort11}, is catastrophic baryon collapse in protogalaxies that 
form in strong LW UV backgrounds at $z \sim$ 15 \citep{wta08,rh09,sbh10,agarw12} 
\citep[see also][]{bl03,jb07b,brmvol08,milos09,awa09,lfh09,th09,li11,pm11,pm12,jlj12a,wf12,
jet13,pm13,latif13a,latif13c,choi13,reis13}.  In this scenario, the primitive galaxy is built up by 
mergers between halos and by accretion in the vicinity of nearby LW UV sources that 
completely suppress star formation in the halos without evaporating them \citep{whb11} 
\citep[see also][about recent numerical models of primeval galaxies]{jgb08,get08,jlj09,get10,
jeon11,pmb11,pmb12,wise12}.  When the galaxy reaches $\sim$ 10$^8$ \Ms, its virial 
temperature crosses the threshold for atomic hydrogen line emission and its baryons begin 
to rapidly cool and collapse. Infall rates at the center of the galaxy can be enormous: 0.1 - 1 
\Ms\ yr$^{-1}$, or 1000 times those in which the first stars form at $z \sim$ 25 \citep{bcl99,
abn00,abn02, bcl02,nu01,on07,on08,wa07,y08,turk09,stacy10,clark11,sm11,get11,get12,
susa13,hir13} \citep[for recent reviews on Pop III star formation, see][]{glov12,dw12}.

\begin{figure*}
\plottwo{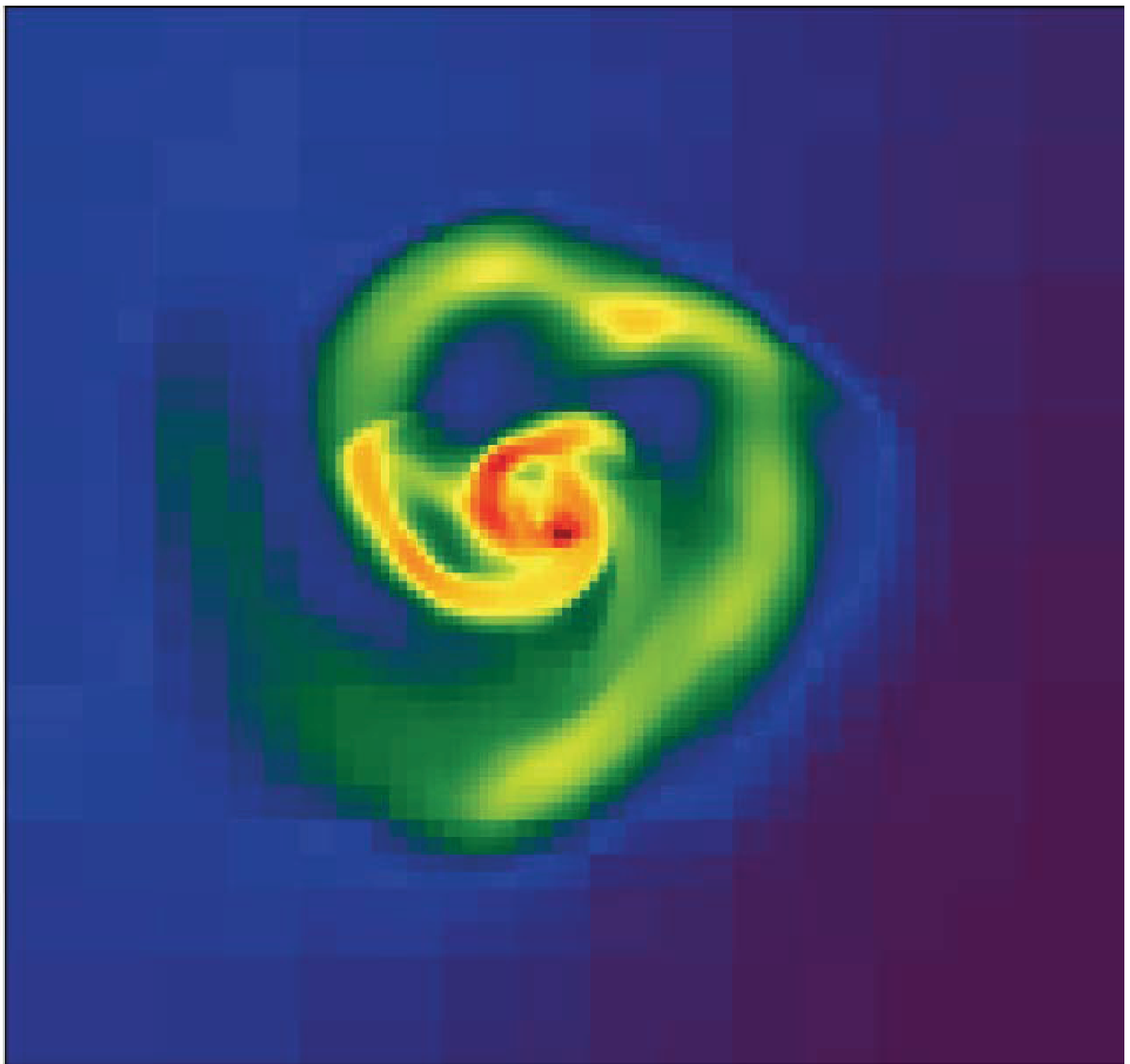}{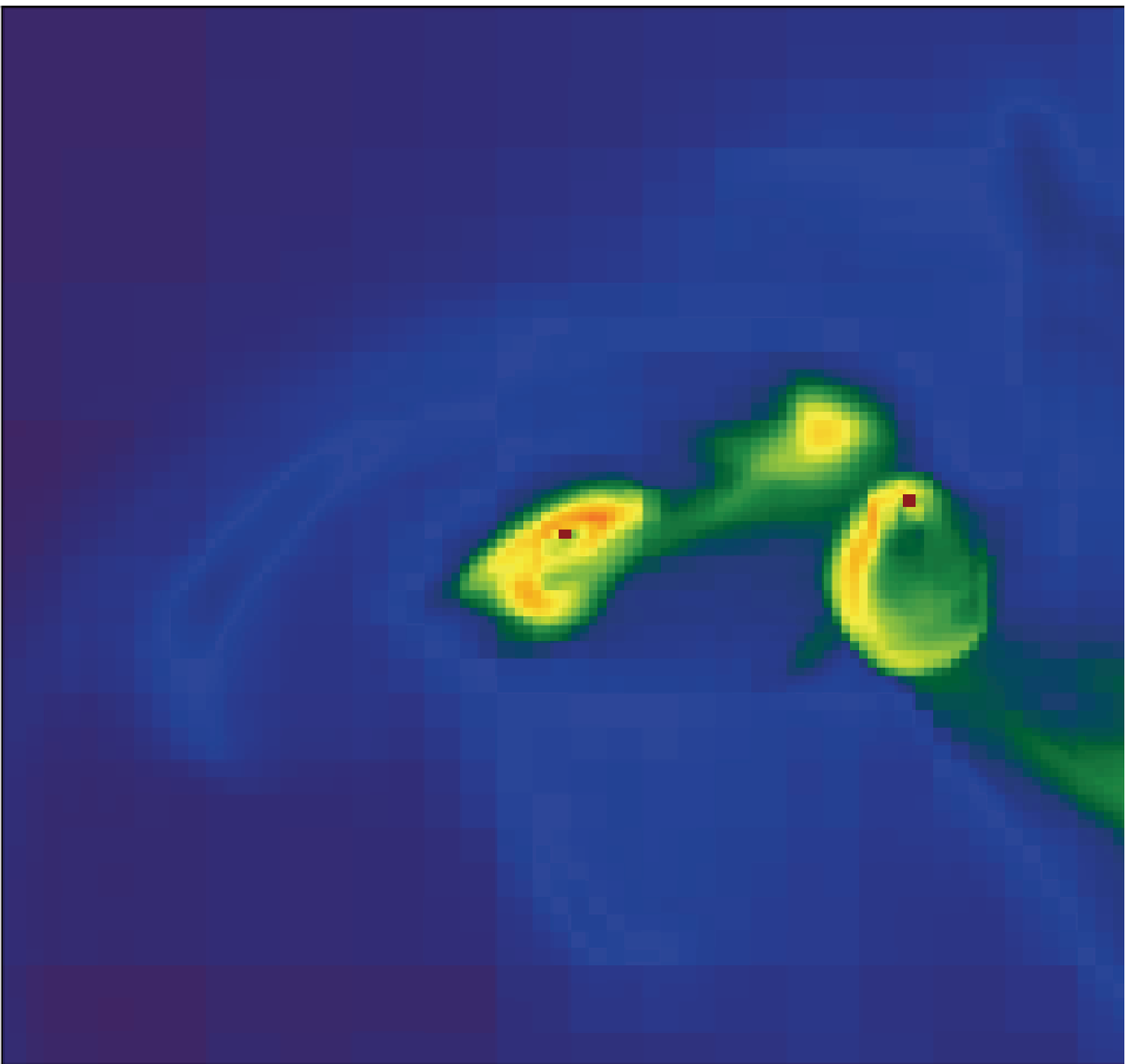}
\caption{Baryon collapse at the center of a $z \sim$ 15 protogalaxy in a LW UV background 
in the \textit{Enzo} AMR code \citep{wet13}.  Left: the formation of an atomically-cooled disk 
at the center of the nascent galaxy at 10,000 yr.  Right:  the breakup of the disk into several 
supermassive fragments shortly thereafter.  The scale is $\sim$ 2000 AU.}
\label{fig:dsk}
\end{figure*}

Numerical simulations show that the baryons can shed angular momentum via the "bars 
within bars" instability on multiple spatial scales and collapse into an isothermal atomically
cooled disk.  The most recent models show that such disks can either feed a single massive 
central object or fragment into several slightly smaller ones \citep{rh09,wet13}.  In Figure 
\ref{fig:dsk} we show the formation and fragmentation of such a disk at the center of a 10$
^8$ \Ms\ halo in a strong LW background at $z \sim$ 15 in an adaptive mesh refinement 
(AMR) simulation done with \textit{Enzo}\footnote{http://code.google.com/p/enzo/}.  The 
recent discovery of a 10$^9$ \Ms\ BH in a quasar at $z \sim$ 7 \citep{mort11} favors SMBH 
seed formation by direct baryon collapse in LW protogalaxies over the creation of BHs by 
Pop III stars at $z \sim$ 25 \citep{wf12,jet13}.

The evolution of the fragments depends on their masses at birth and subsequent accretion 
histories \citep[see][for studies of Pop III stellar evolution under ongoing accretion at much 
lower rates]{ohk09}.  One possibility is that the fragment forms a supermassive star \citep{
fuller86}.  In most cases these stars collapse directly to black holes.  Another possibility is 
that the core of the star becomes a black hole whose radiation supports the upper, 
convective layers of the star against collapse. The result is a "quasistar" that appears to an 
external observer to be a large, cool star that is powered by a black hole at its center rather 
than by a fusion core \citep{begel06,begel08,begel10,vb10}.  There is some question as to 
whether a quasistar could be stable because because a jet from the BH could rupture the 
upper layers of the star and destabilize its envelope, but the final result would be the same:  
a 10$^4$ - 10$^5$ \Ms\ SMBH seed.

The fragment could instead collapse quasistatically with intermittent nuclear burning 
without ever entering the main sequence. If it has enough angular momentum a black  
hole accretion disk (BHAD) system might form form at its center.  In this case most of 
the fragment eventually falls into the BH, perhaps with the formation of a strong wind 
that drives nuclear burning in the disk and blows some heavy elements out of the clump.  
However, if infall onto the fragment is heavy during its collapse it may be difficult to form 
a standard, geometrically thin accretion disk because this would require photon leakage 
times (radiative cooling times) to be shorter than the accretion (inflow) times, an ordering 
of time scales that is unlikely under these circumstances.  

If a supermassive star forms in these conditions its radiation may not halt accretion and 
it too may evolve under heavy infall over its entire life \citep{jlj12a}.  In contrast, 
lower-mass Pop III stars usually disperse baryons from their halos \citep{wan04,ket04,
abs06,awb07,wa08a,wn08a,wn08b}. Other mechanisms for massive fragmentation and 
SMBH seed formation have been proposed, such as the suppression of gas cooling by 
primordial magnetic fields \citep{sethi10} and cold accretion shocks \citep{io12}.

We have found that for a narrow range of mass around 55,000 \Ms, atomically-cooled 
fragments can settle into stable nuclear burning and become supermassive stars with
lifetimes of $\sim$ 2 Myr \citep{heg13}.  These stars die as extremely energetic 
thermonuclear SNe, with energies of $\sim$ 10$^{55}$ erg, or 100 times those of 65 - 
260 \Ms\ Pop III PI SNe \citep{rs67,brk67,hw02,byh03,ky05,gy09,ds11a,vas12} \citep[see 
also][]{montero12}.  Such events would be the most energetic explosions in the cosmos, 
and their detection could reveal the birthplaces of SMBHs created by direct collapse, 
since LW protogalaxies are the only environments known to form such massive clumps 
at $z \sim$ 15. Could such SNe be discovered by existing or future observatories? 
\citet{wet12a,wet12b} recently found that 140 - 260 \Ms\ Pop III PI SNe will be detected 
in the NIR out to $z \ga$ 30 by the \textit{James Webb Space Telescope} 
\citep[(\textit{JWST})][]{jwst06} and to $z \sim$ 15 - 20 in all-sky surveys by the 
\textit{Wide Field Infrared Survey Telescope} (\textit{WFIRST}) and the \textit{Wide 
Field Imaging Surveyor for High Redshift} (\textit{WISH}) \citep[see also][]{wa05,sc05,
fwf10,kasen11,hum12,pan12a,pan12b,wet12c,wet12e,ds13}.   However, supermassive 
SNe might occur in very dense accretion envelopes that quench their luminosities at 
early times, when the greatest fraction of their flux is redshifted into the NIR.  It is not 
clear if explosions could be detected.

We present numerical simulations of light curves, spectra and NIR signals of 55,500 \Ms\ 
Pop III SNe at 7 $<z< $ 30 done with the Los Alamos RAGE and SPECTRUM codes. We 
consider only the observational signatures of these events, and defer detailed discussion 
of progenitor evolution, explosive nucleosynthesis and multidimensional mixing to \citet{
heg13}. The effects of these explosions on the protogalaxies that host them are examined 
in \citet{jet13a}, \citet{wet13a}, and \citet{wet13b}.  In Section 2 we review our numerical 
methods for evolving the star, its explosion, and the propagation of the blast through the 
star and its envelope.  In Section 3 we examine blast profiles, light curves and spectra for 
the SN in the source frame, and in Section 4 we show its NIR light curves in the observer 
frame and calculate detection thresholds as a function of redshift for these explosions.  In 
Section 5 we conclude by discussing complementary detection strategies for the formation
of SMBH seeds via direct collapse.

\section{Numerical Method}

We calculate light curves and spectra in three stages.  First, we evolve the 55,500 \Ms\ 
zero-metallicity star from the beginning of the main sequence through collapse, explosive 
nuclear burning, and the expansion of the shock to the edge of the star in the Kepler code.  
In a parallel calculation we map the Kepler profile of the star onto a two-dimensional (2D) 
AMR mesh in the CASTRO code and evolve it through the same stages as in our Kepler 
model.  Next, we spherically average mass fractions from the final CASTRO profile onto a 
1D spherical-coordinate grid in RAGE along with final density, velocity and energy profiles 
from the Kepler calculation.  This is done to approximate how mixing in the interior of the 
star prior to shock breakout affects explosion spectra at later times.  We evolve the shock 
through the surface of the star and into the surrounding medium with RAGE until it dims 
below observability. Finally, we post process our RAGE profiles with the SPECTRUM code 
to calculate light curves and spectra. 

\subsection{Kepler}

We determine the internal structure of the star at the time of the explosion by evolving it 
from the beginning of the main sequence to the onset of collapse in the one-dimensional
(1D) stellar evolution code Kepler \citep{Weaver1978,Woosley2002}. The SN begins 
when the core of the star begins to contract and initiates explosive burning in the O and 
Si layers \citep{hw02,hw10,jw11,chen11}.  The non-rotating star, which is resolved with 
1148 mass zones, lives for 1.69 Myr and then dies as blue giant with a radius of 1.33 $
\times$ 10$^{13}$ cm, similar to those of the z-series stars in \citet{wet12b} even though 
they are $\sim$ 200 times as massive.  The mass of the He core at the time of the 
explosion is 2.67 $\times$ 10$^{4}$ \Ms.  

This treatment is approximate for several reasons.  First, we do not model the pre-main 
sequence evolution of the star or its growth from much lower masses.  Instead, the star 
is initialized at the beginning of the main sequence in our evolution calculations.  Second, 
we exclude the ongoing accretion under which the star may evolve over its lifetime, which 
might alter its final properties.  Third, we do not include stellar rotation, which could lower 
the mass at which the star explodes \citep{cw12,cwc13}.  Rotation may also broaden the 
mass range for which supermassive fragments can actually become stars by supporting 
them against collapse and enabling stable nuclear burning.  Finally, we do not include 
radiative feedback from the star on inflow, which could regulate its growth rates. However, 
in some cases luminosity from the star terminates accretion \citep{jlj12a} so our 
assumption that the star has a constant mass would be valid.  The evolution of massive 
primordial clumps and supermassive stars under ongoing accretion will be the focus of 
future studies.

\subsection{CASTRO}

\begin{figure*}
\begin{center}
\begin{tabular}{ccc}
\epsfig{file=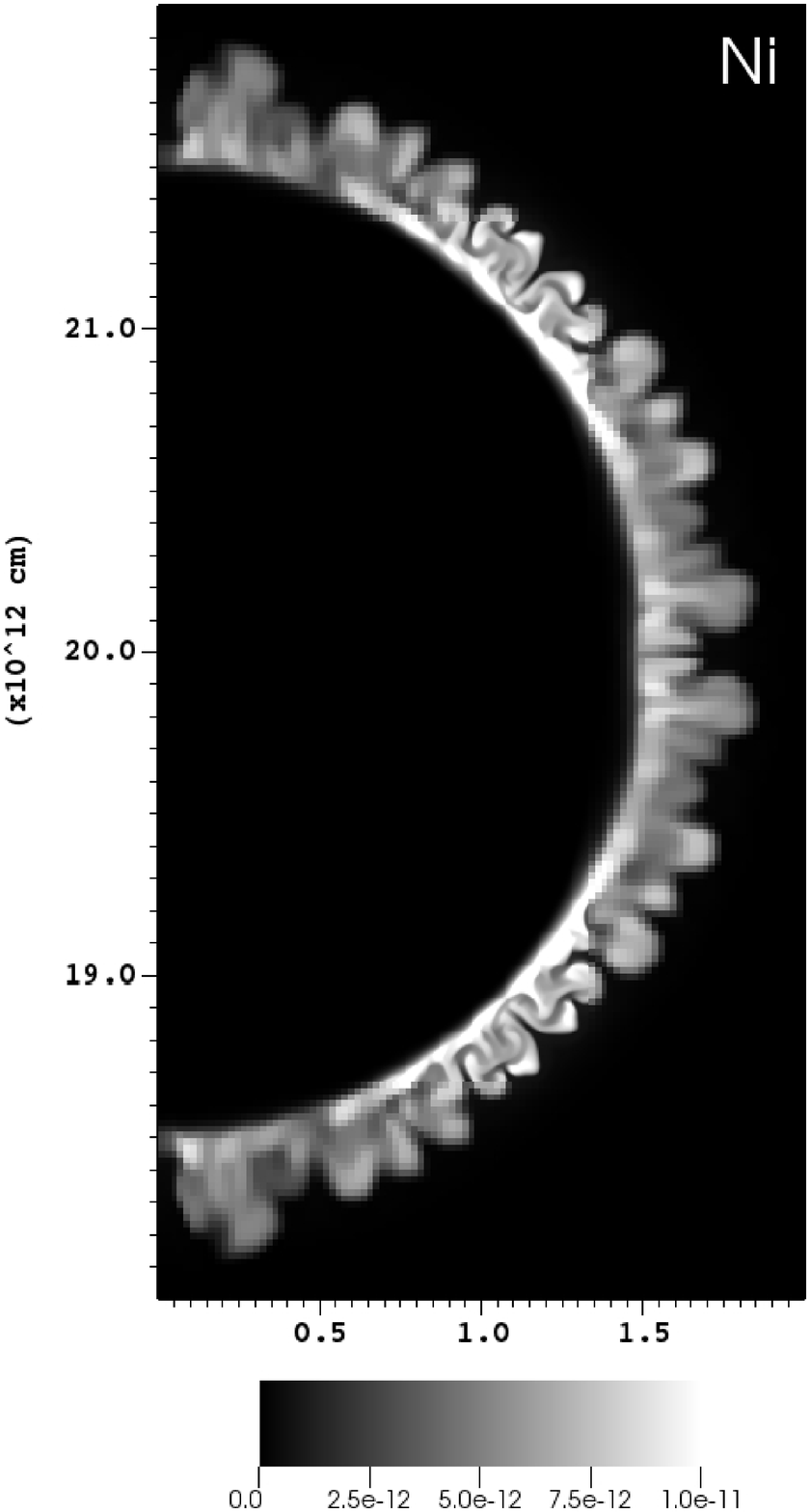,width=0.3\linewidth,clip=} & 
\epsfig{file=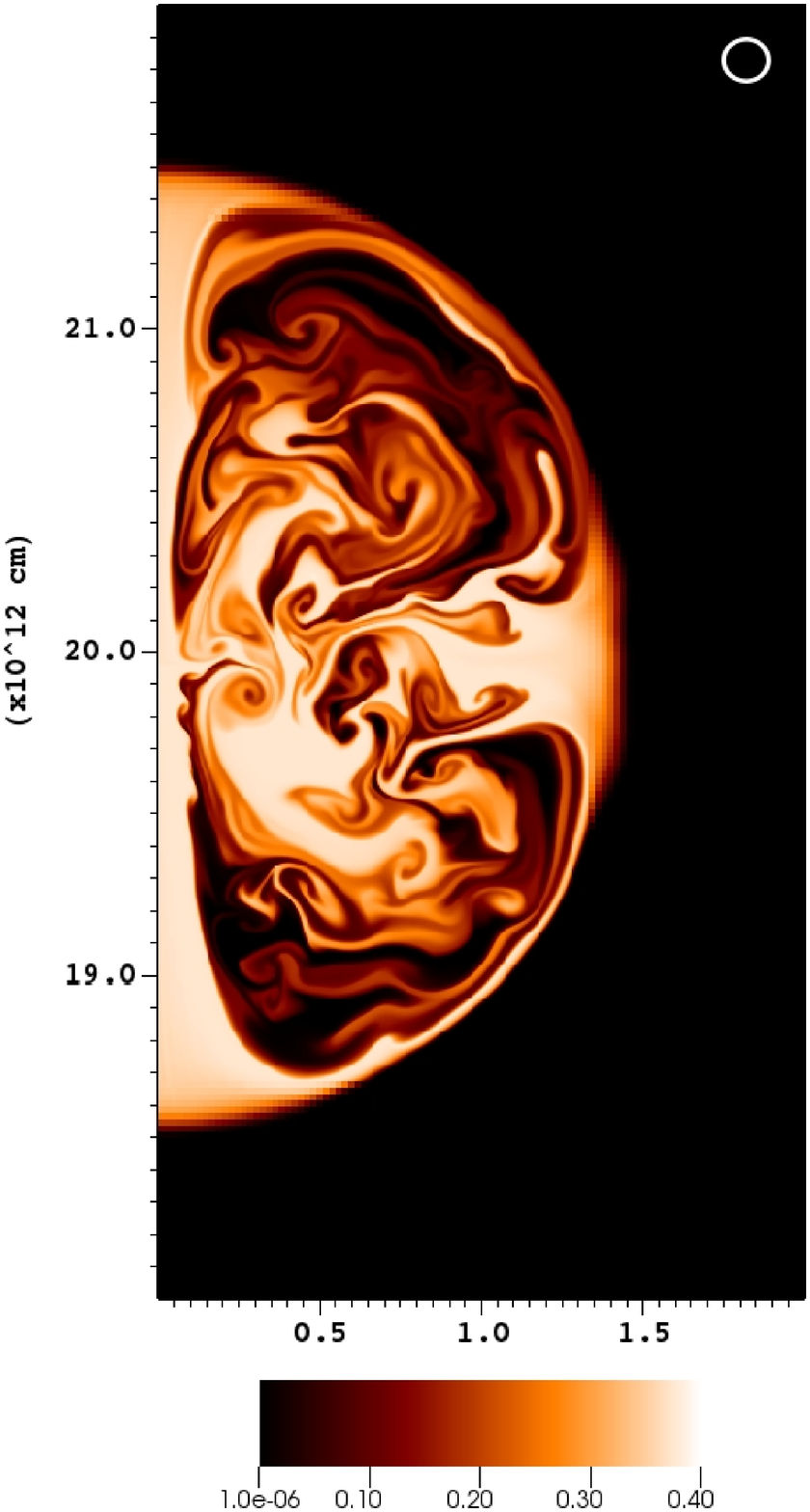,width=0.295\linewidth,clip=} &
\epsfig{file=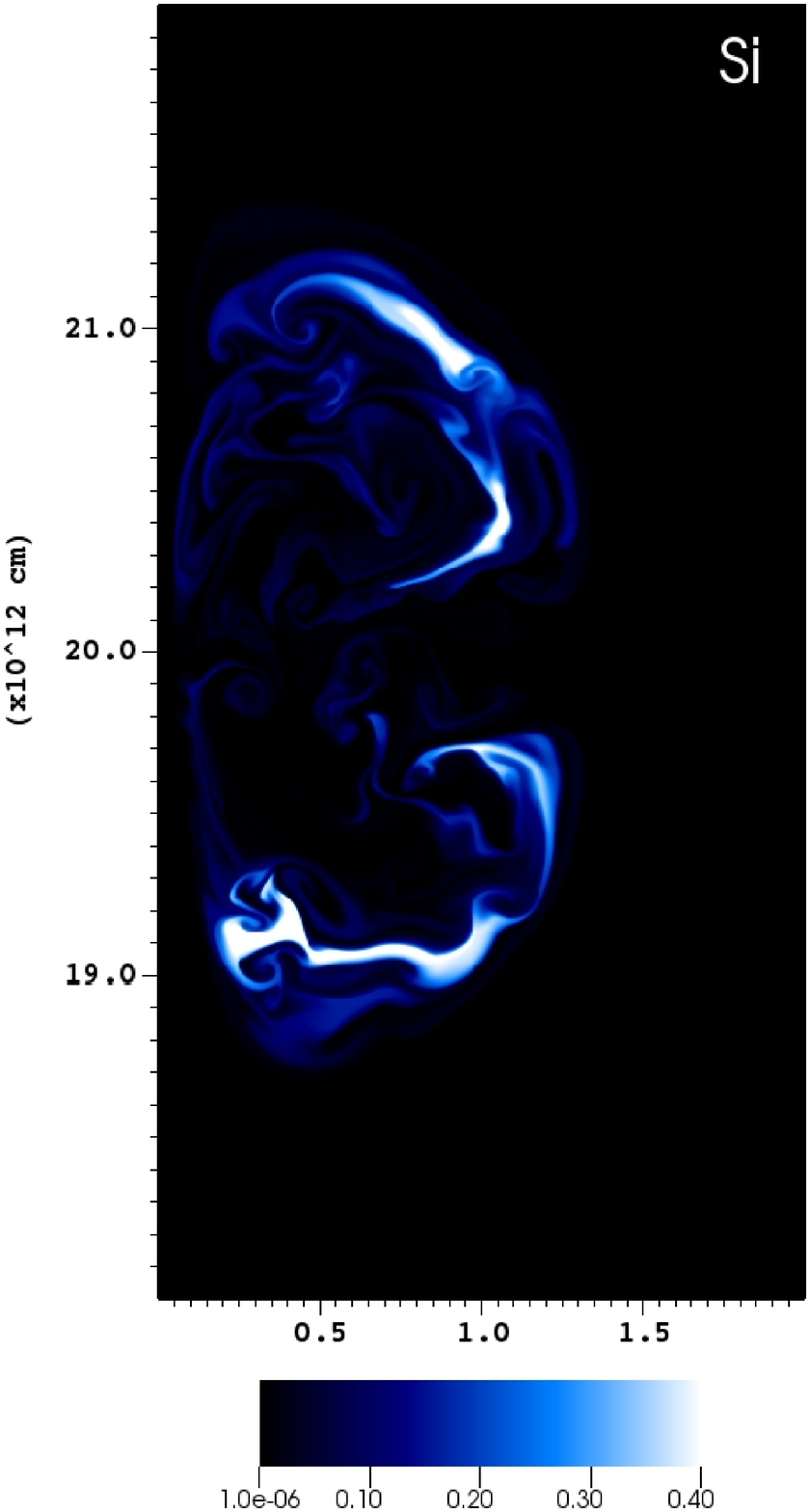,width=0.3\linewidth,clip=} 
\end{tabular}
\end{center}
\caption{Mixing in the Ni shell (left), O shell (center) and Si shell (right) just before shock
breakout in CASTRO.  The images are mass fractions.}
\label{fig:mixing}
\end{figure*}

At the beginning of central collapse we map our Kepler profiles onto a 2D axisymmetric 
grid in CASTRO \citep{Almgren2010} and then evolve the SN through collapse and 
explosive burning, halting the simulation when the shock reaches the edge of the star.  
CASTRO (Compressible ASTROphysics) is a multidimensional Eulerian AMR code with 
an unsplit Godunov hydrodynamics solver.  Energy production is calculated with a 
19-isotope network up to the point of oxygen depletion in the core and with a 128-isotope 
quasi-equilibrium network thereafter.  We evolve mass fractions for the same 15 even
numbered elements that are predominantly synthesized by PI SNe.  Radiation transport 
is not required in these models because the mean free paths of photons prior to breakout 
are so short that they are simply advected through the star by the fluid flow.  We include 
the contribution of photons to the gas pressure in the equation of state.  Our models 
include energy deposition due to radioactive decay of \Ni\ in the ejecta as described by 
equation 4 in \citet{jet09b} although, as we discuss below, this explosion produces very 
little \Ni, unlike 140 - 260 \Ms\ Pop III PI SNe.

Mapping an explosion profile from a 1D Lagrangian coordinate mesh in mass to a 2D 
mesh in space can lead to violations in conservation of mass and energy.  Linear 
interpolations in radius can also fail to resolve key features of the original profile, such 
as the structure of the core of the star and its temperature profile.  Failure to properly 
map temperature features can be especially problematic because nuclear burn rates are 
highly sensitive to them during the explosion.  To avoid these difficulties, we port Kepler 
profiles to CASTRO with the new conservative mapping scheme of \citet{chen11}.  This 
approach conserves mass and energy while reproducing all the features of the original 
profile over a broad dynamical range in space.  The CASTRO root grid is 256$^2$ with 
a resolution of 2.0 $\times$ 10$^{10}$ cm and up to two levels of AMR refinement (a 
factor of four increase in resolution).

\begin{figure*}
\plottwo{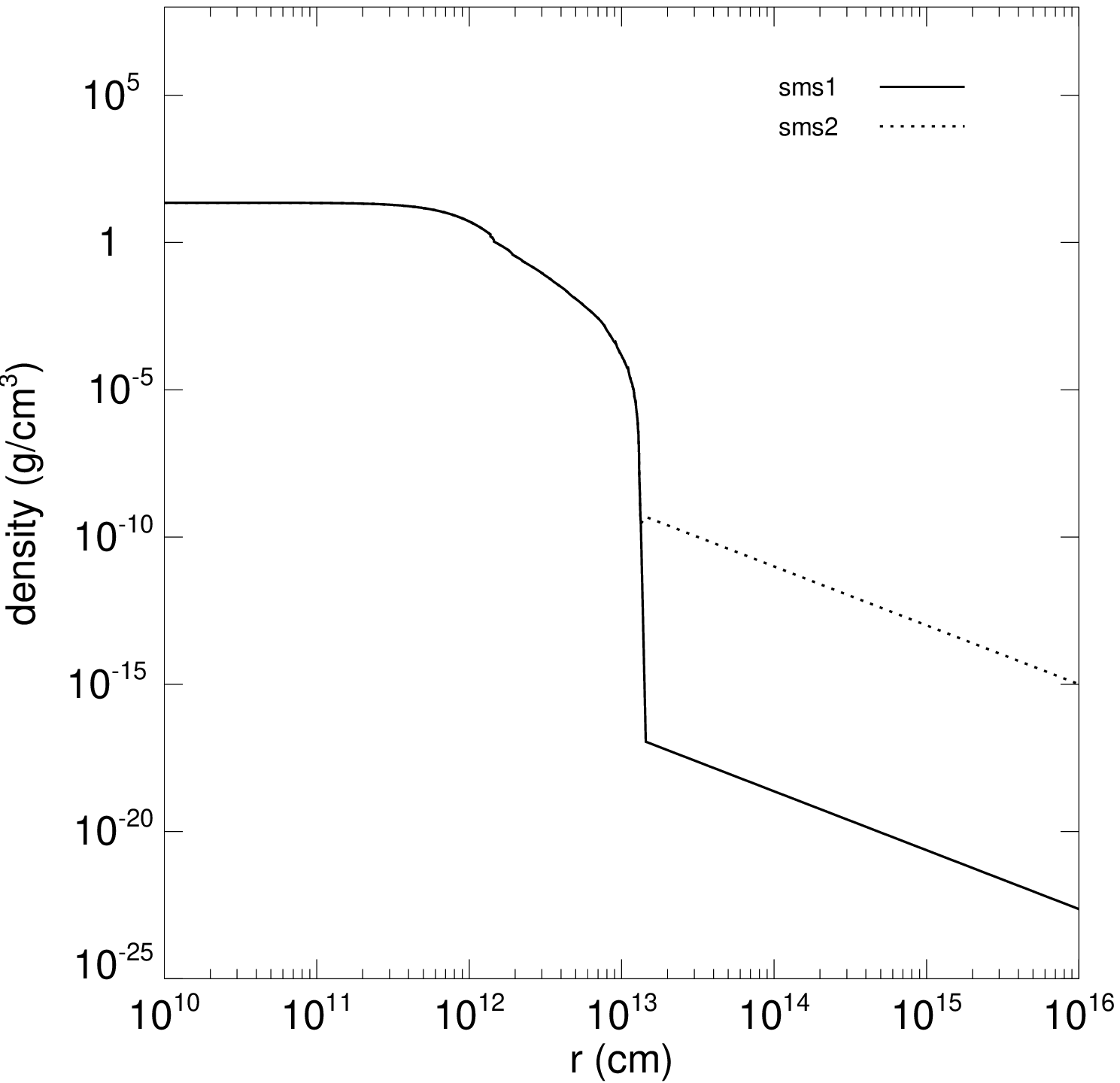}{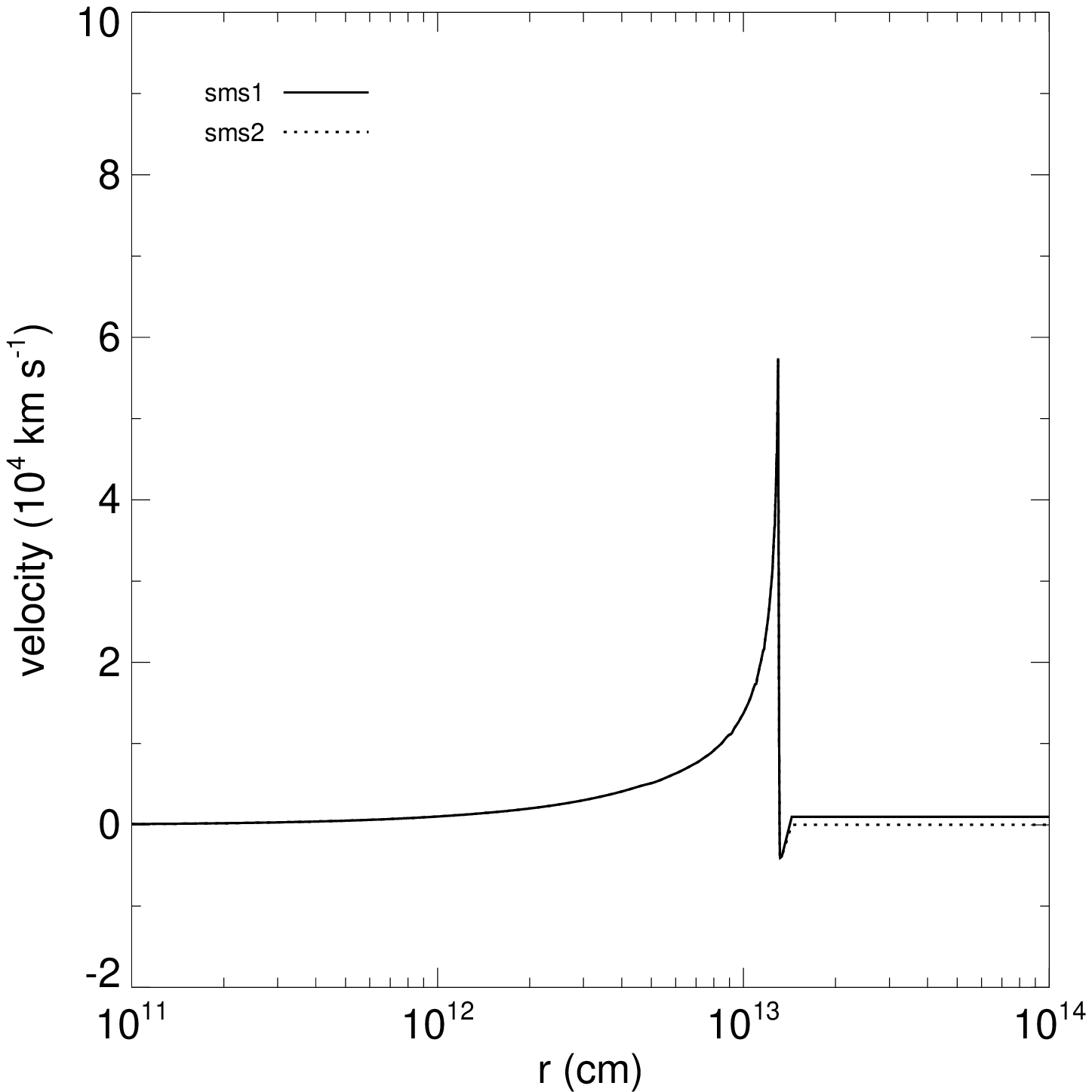}
\caption{Initial RAGE explosion profiles.  Right:  densities.  Left:  velocities} \vspace{0.1in}
\label{fig:initprof}
\end{figure*}

The star explodes with an energy of 7.74 $\times$10$^{54}$ erg.  Explosive burning begins 
in the O and Si layers and is done by $\sim$ 100 s.  The SN creates only trace amounts of 
\Ni, $\sim$ 2.25 $\times$ 10$^{-8}$ \Ms, unlike Pop III PI SNe that form up to 50 \Ms\ of \Ni.  
The core of the SMS does not burn all the way to \Ni\ like in PI SNe, and the little that is 
formed is at the edge of the He layer.  As shown in Figure \ref{fig:mixing}, the shock heavily 
mixes the interior of the star by the time it reaches the surface, in contrast to Pop III PI SNe 
that exhibit little mixing \citep{jw11}. The mixing is driven by fluid instabilities that are seeded 
during collapse and then amplified by explosive burning rather than by the formation of a 
reverse shock and the subsequent appearance of Rayleigh-Taylor instabilities at later times, 
as in 15 - 40 \Ms\ Pop III SNe \citep{jet09b}.  Mixing is important to SN spectra because it 
can determine the order in which emission and absorption lines appear over time.  Mass 
fractions for the various elements are realistically distributed in radius and angle in CASTRO 
when the shock breaks out of the star.  Spherically averaging them prior to mapping them 
into RAGE therefore allows us to capture how mixing governs the order in which lines later
appear in the spectra over time even though our RAGE models are 1D.  We halt the 
CASTRO run when the shock is $\sim$ 100 photon mean free paths $\lambda_{\mathrm{p}}$ 
from the edge of the star: 
\vspace{0.05in}
\begin{equation}
\lambda_{\mathrm{p}} = \frac{1}{\kappa_{\mathrm{Th}}\rho}, \vspace{0.05in}
\end{equation}
where $\kappa_{\mathrm{Th}}$ is the opacity due to Thomson scattering from electrons 
(0.288 gm$^{-1}$ cm$^2$ for primordial gas) and $\rho$ is the density just beyond the 
shock inside the star. 

\subsection{RAGE}

We evolve the shock through the surface of the star and its envelope with the Los Alamos 
National Laboratory (LANL) radiation hydrodynamics code RAGE \citep{rage}.  RAGE 
(Radiation Adaptive Grid Eulerian) is a multidimensional AMR code that couples second
order conservative Godunov hydrodynamics to grey or multigroup flux-limited diffusion 
(FLD) to model strongly radiating flows.  RAGE utilizes the LANL OPLIB atomic opacity 
database\footnote{http://aphysics2/www.t4.lanl.gov/cgi-bin/opacity/tops.pl} \citep{oplib} 
and can evolve multimaterial flows with a variety of equations of state (EOS).  We employ 
the same physics as in \citet{fet12}: multispecies advection, grey FLD radiation transport 
with 2-temperature (2T) physics and LTE opacities, energy deposition from the radioactive 
decay of \Ni, and an ideal gas EOS.  2T physics better captures shock breakout, when 
radiation and matter temperatures can be out of equilibrium.  We advect mass fractions for 
15 elements, the even numbered elements predominantly synthesized by PI SNe.  

\begin{figure}
\plotone{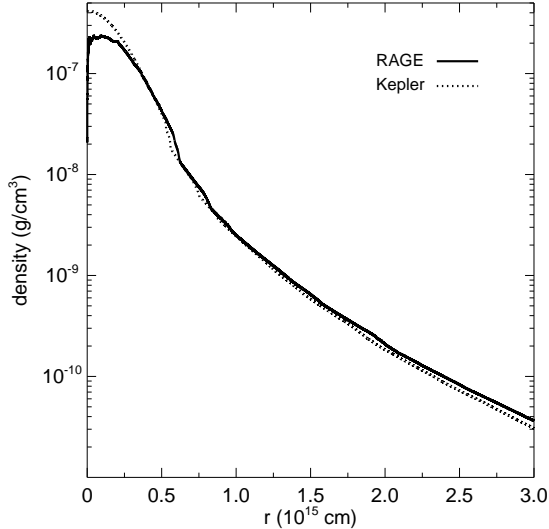}
\caption{Density profiles for the supermassive SN at 2.9 $\times$ 10$^9$ s for RAGE 
and Kepler.}
\label{fig:selfgrav}
\end{figure}

As in \citet{wet12b}, we include the self-gravity of the ejecta in our simulations.  Because
so much mass is packed into such a small volume in the star, its initial potential energy is 
close to the energy released in the explosion and must be included to obtain the kinetic 
energy and luminosity of the shock at early times.  As a test of our recent implementation 
of self-gravity in RAGE we evolved the SN from just before shock breakout to 2.9 $\times$ 
10$^6$ s in both RAGE and Kepler.  As we show in Fig. \ref{fig:selfgrav}, the two density 
profiles are essentially identical at the latter time.  The minor differences at the center are 
attributable to differences between the hydrodynamics schemes of the two codes.

We spherically average densities, velocities, specific internal energies (erg gm$^{-1}$), 
and species mass fractions from CASTRO onto a 200,000 zone 1D spherical mesh in 
RAGE.  Since radiation energy densities are not explicitly evolved in Kepler, we initialize 
them in RAGE as 
\vspace{0.05in}
\begin{equation}
E_{rad} = aT^4, \vspace{0.05in}
\end{equation}
where $a =$ 7.564 $\times$ 10$^{-15}$ erg cm$^{-3}$ K$^{-4}$ is the radiation constant 
and $T$ is the gas temperature.  Also, because gas energies in Kepler include 
contributions from the ionization states of atoms, we construct the specific internal energy 
from $T$ with
\vspace{0.05in}
\begin{equation}
E_{gas} = C_VT, \vspace{0.05in}
\end{equation}
where $C_V = $ 1.2472 $\times$ 10$^{8}$ erg K$^{-1}$ is the specific heat of the gas.  

At the beginning of the simulation we resolve the region from the center of the grid to the 
edge of the shock in the velocity profile with 100,000 zones.  We allow up to five levels of 
refinement during the initial mapping of the profile but turn off AMR during the simulation.  
Our grid ensures that the photosphere of the shock is always resolved since failure to do 
so can lead to underestimates of luminosity during post processing.  The radius of the 
shock in our setup is 1.3 $\times$ 10$^{13}$ cm and our first grid has a resolution of 1.3 
$\times$ 10$^8$ cm with an outer boundary at 2.6 $\times$ 10$^{13}$ cm.

We set reflecting and outflow boundary conditions on the fluid and radiation variables at 
the inner and outer boundaries of the grid, respectively. At the beginning of the simulation,
Courant times are short due to high temperatures, large velocities and small cell sizes. To 
minimize execution times and to accommodate the expansion of the SN, we periodically 
regrid the profiles onto a larger mesh as the explosion grows.  At each regrid we allocate 
100,000 zones out to either the edge of the shock (pre-breakout) or the radiation front 
(post-breakout).  In the latter case we take the radius at which the radiation temperature 
falls to the wind temperature (0.01 eV) to be the edge of the front.  The inner boundary is 
always at the origin and the outer boundary of the final, largest mesh in our simulations is 
1.0 $\times$  10$^{18}$ cm.  We again permit up to five levels of refinement during the 
initial regridding of the profile but disable AMR during the simulation.

\subsection{Circumstellar Envelope}

We consider explosions in two kinds of envelope: low-mass outflows (SMS1) and massive 
inflows like those that grew the star to such large masses in such short times (SMS2). For 
diffuse outflows we adopt the usual power-law density profile for a wind at a constant 
velocity:
\vspace{0.1in}
\begin{equation}
\rho_{\mathrm{W}}(r) = \frac{\dot{m}}{4 \pi r^2 v_{\mathrm{W}}}. \vspace{0.1in}
\end{equation}
Here, $\dot{m}$ is the mass loss rate associated with the wind and $v_{\mathrm{W}}$ is 
the wind speed. The mass loss rate is calculated from the total mass loss $M_{\mathrm{
tot}}$ and the main sequence lifetime of the star $t_{\mathrm{MSL}}$: \vspace{0.1in}
\begin{equation}
\dot{m} = \frac{M_{\mathrm{tot}}}{t_{\mathrm{MSL}}}. \vspace{0.1in} \label{eq:wind}
\end{equation}
Pop III stars are not thought to lose much mass over their lives because there are no 
metals in their atmospheres to drive winds \citep{Kudritzki00,Baraffe01,Vink01,kk06}, 
so we set $M_{\mathrm{tot}} =$ 0.1 \Ms\ and $v_{\mathrm{w}} =$ 1000 km s$^{-1}$. 

We treat the massive infall envelope as a wind in reverse, with $\dot{m}$ = 0.01 \Ms\ yr$
^{-1}$ and an infall velocity $v_{\mathrm{w}} =$ 5 km s$^{-1}$, in keeping with numerical
simulations of baryon collapse in protogalaxies in strong LW backgrounds.  This profile 
assumes that accretion is spherical when in reality it may occur in a disk, so it should be 
considered to be the densest envelope through which the SN shock might propagate.  In 
both cases we take the wind to be 76\% H and 24\% He by mass.  Rather than calculate 
the ionization state of the wind \citep{wn06} we take it to be cold ($T =$ 0.01 eV) and 
neutral in all our models for simplicity. This assumption holds for dense envelopes, where 
semi-analytical studies have shown that ionizing UV photons cannot propagate more than 
a few dozen stellar radii from the star over its lifetime \citep{jlj12a} \citep[for studies on UV 
breakout from low-mass Pop III protostellar disks, see][]{op01,oi02,op03,tm04,tm08,hos11,
stacy12,hos12}. 

On the other hand, the compact blue progenitor, with a total luminosity of 3.5 $\times$ 
10$^{42}$ erg s$^{-1}$ and $T_{\mathrm{eff}}$ = 6.85 $\times$ 10$^4$ K, likely ionizes 
the diffuse wind so the luminosities we calculate for that case are lower limits.  We show 
initial RAGE density and velocity profiles for the shock, the star, and its envelope in Fig. 
\ref{fig:initprof}.  The surface of the star is visible as the sharp drop in density at $\sim$ 
1.4 $\times$ 10$^{13}$ cm.  The fact that the accretion envelope has a density at the 
surface of the star that is seven orders of magnitude greater than that of the wind has 
important consequences for shock and radiation breakout, as we discuss below.  In both 
cases we evolve the SN out to 3 yr.

\subsection{SPECTRUM}

\begin{figure*}
\plottwo{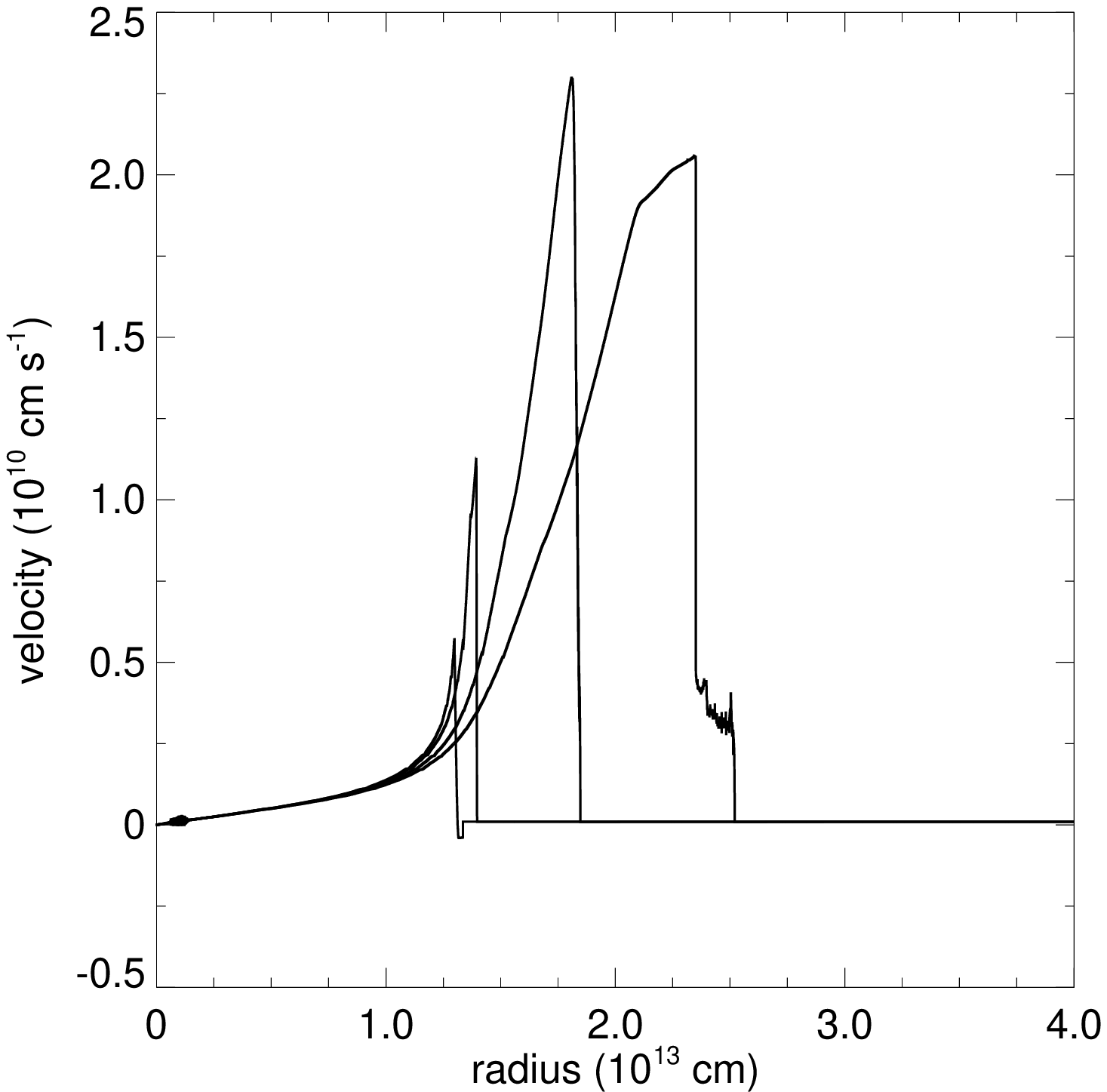}{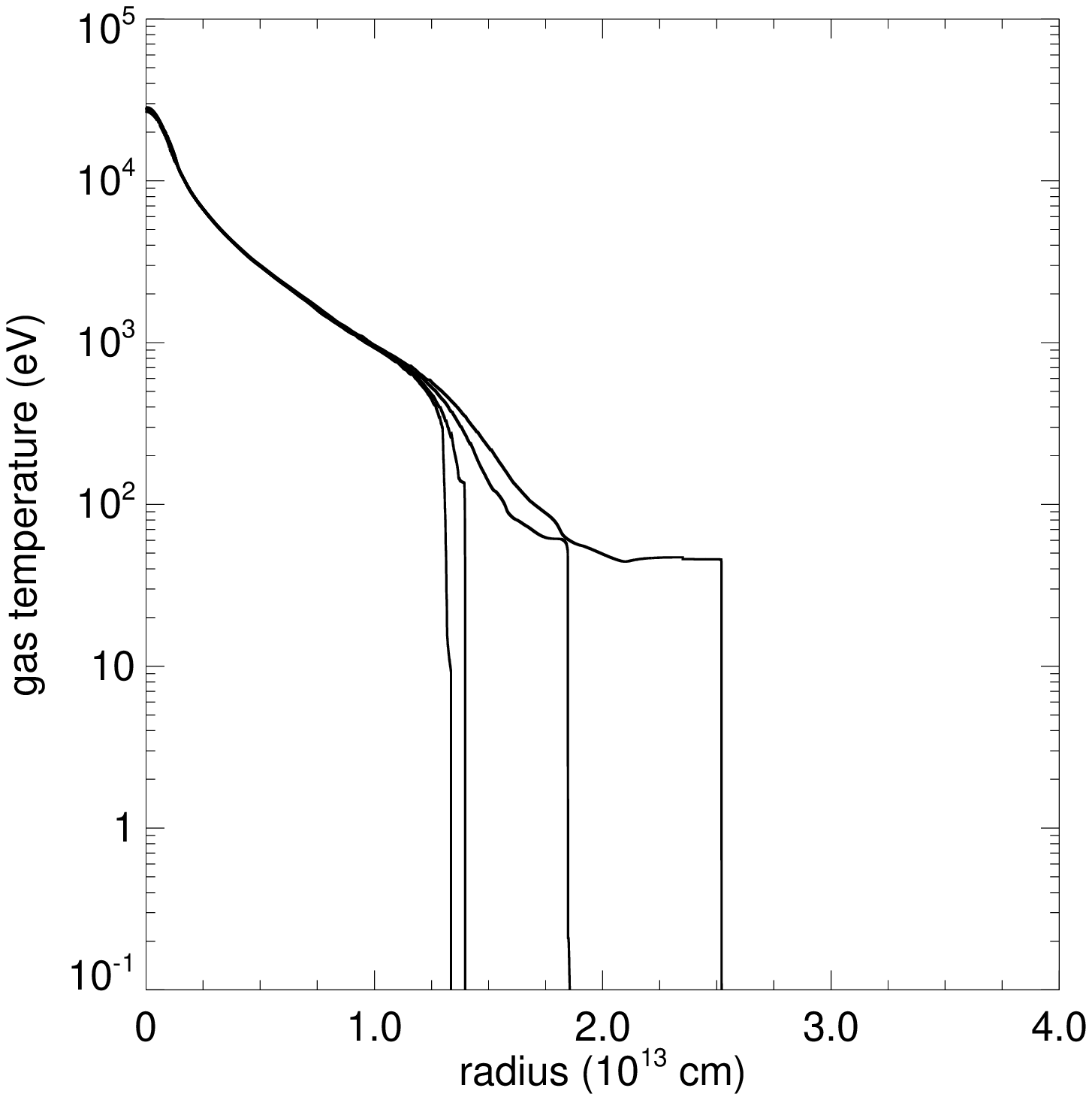}
\caption{Shock breakout into a diffuse envelope (SMS1).  Right:  velocities; from left to 
right:  7900 s, 7987 s, 8192 s and 8407 s.  Left:  temperatures at the same times.}
\label{fig:sbo1}
\end{figure*}

\begin{figure*}
\plottwo{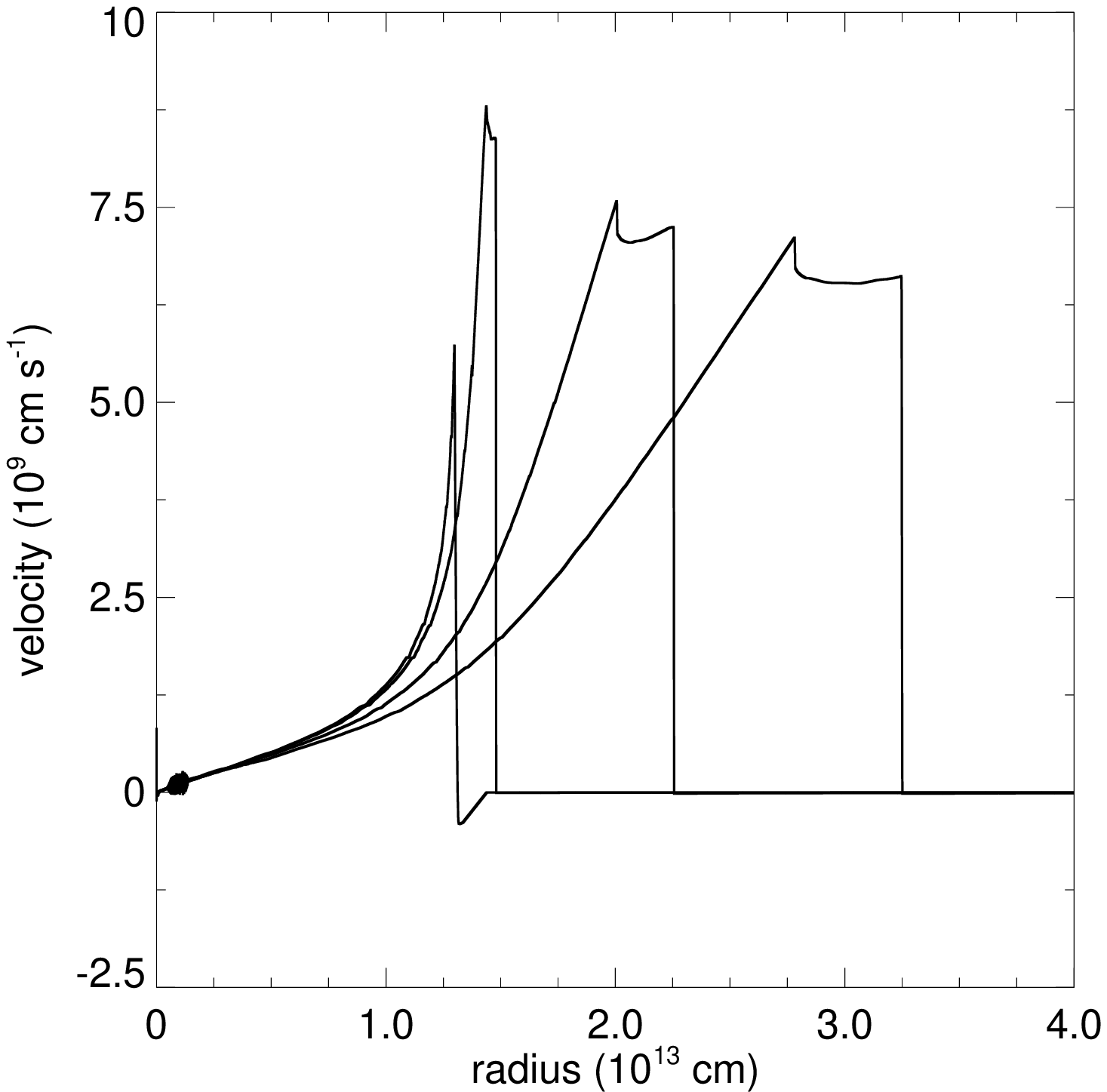}{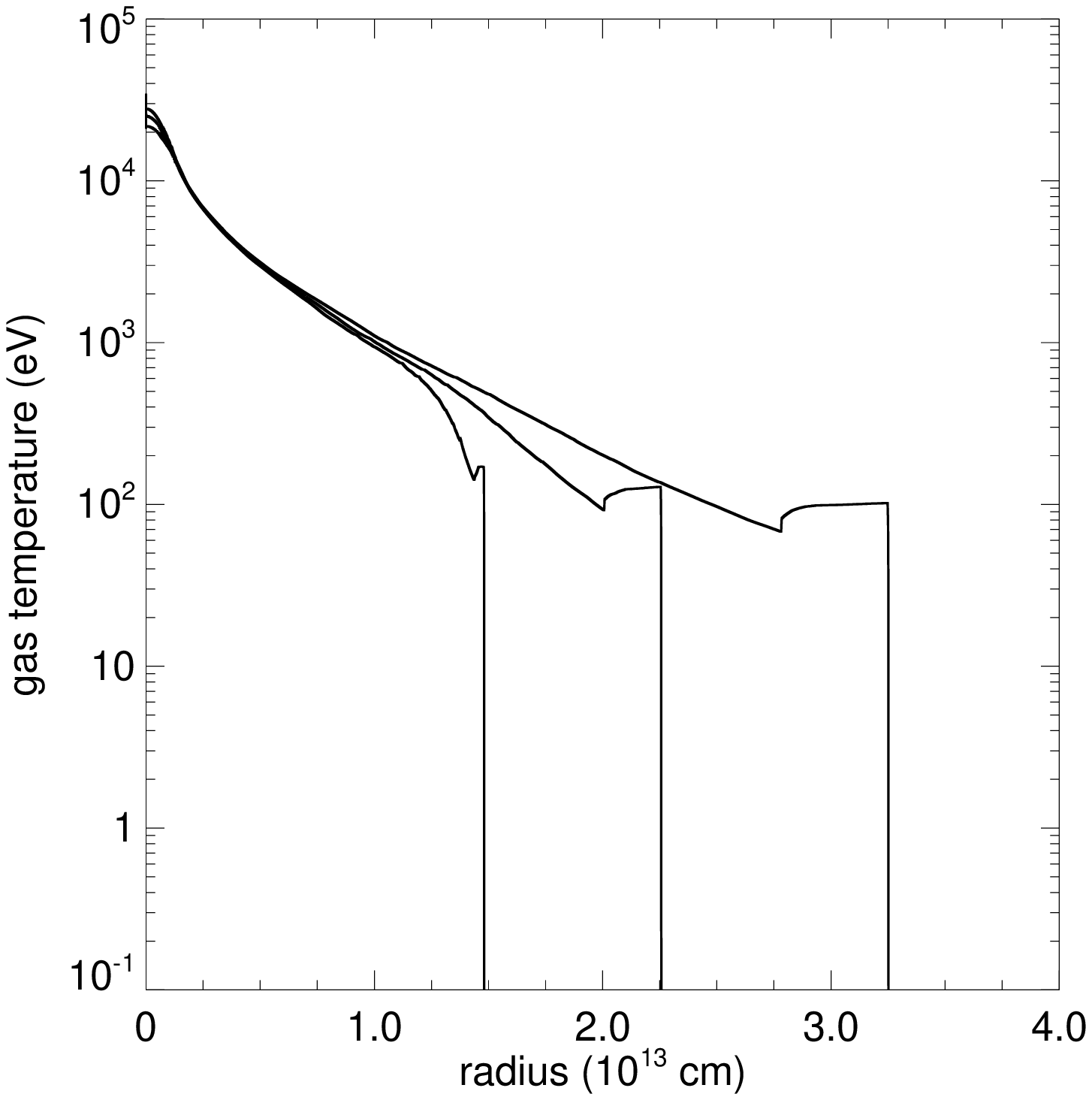}
\caption{Shock breakout into dense infall (SMS2).  Right:  velocities; from left to right:  
7900 s, 8083 s, 8940 s and 1.02e4 s.  Left:  temperatures at the same times.}
\label{fig:sbo2}
\end{figure*}

We calculate spectra for the explosions with the LANL SPECTRUM code.  SPECTRUM 
directly sums the luminosity of every fluid element in a SN profile to calculate the total 
flux escaping the ejecta along the line of sight for 14900 wavelengths.  The procedure 
is described in detail in \citet{fet12} and accounts for Doppler shifts and time dilation due 
to the relativistic expansion of the ejecta.  SPECTRUM also calculates the intensities of 
emission lines and the attenuation of flux along the line of sight with OPLIB opacities, 
so it captures limb darkening and absorption lines imprinted on the flux by intervening 
material in the SN ejecta and envelope.

As explained in \citet{fet12}, densities, velocities, radiation temperatures and mass 
fractions from the finest levels of refinement in the RAGE AMR hierarchy are extracted 
and ordered by radius into separate files, with one variable per file.  These profiles can 
contain more than 200,000 radial zones, so limits on machine memory and time prevent 
us from using all of them to calculate a spectrum.  We therefore map only a subset of 
the points onto the new grid.  We first sample the radiation energy density profile inward 
from the outer boundary to find the position of the radiation front, which we define to be 
where $aT^4$ rises above 1.0 erg/cm$^3$.  This energy density is intermediate to that 
of the cold wind and the radiation front.  The radius of the $\tau = $ 25 surface is then 
found by integrating the optical depth due to Thomson scattering inward from the outer 
boundary, where $\kappa_{Th} =$ 0.288 gm$^{-1}$ cm$^2$ for primordial H and He.  
This gives the greatest depth from which photons can escape from the ejecta because 
$\kappa_{Th}$ is the minimum total opacity.  

To compute a spectrum, we interpolate the densities, temperatures, velocities and 
mass fractions we extract from RAGE onto a 2D grid in $r$ and $\mu =$ cos$\, 
\theta$ in SPECTRUM, whose inner and outer boundaries are zero and 10$^{18}$ cm.  
The region from the center of the grid to the $\tau =$ 25 surface is partitioned into 800
uniform zones in log radius.  We allocate 6200 uniform zones in radius between the 
$\tau =$ 25 surface and the edge of the radiation front.  The wind between the front 
and the outer boundary is divided into 500 uniform zones in log radius, for a total of 
7500 radial bins. The fluid variables in each of these new radial bins is mass averaged 
to ensure that SPECTRUM captures very sharp features in the original RAGE profile.  
The grid is discretized into 160 uniform zones in $\mu$ from -1 to 1.  Our choice of 
mesh yields good convergence in spectrum tests, fully resolving regions of the flow 
from which photons can escape the ejecta and only lightly sampling those from which 
they cannot.

\section{Blast Profiles, Light Curves and Spectra}

We show velocity and gas temperature profiles at shock breakout for the SMS1 and
SMS2 explosions in Figures \ref{fig:sbo1} and \ref{fig:sbo2}. Before breakout, the SN 
cannot be seen by an external observer because photons from the shock are 
scattered by e$^-$ in the upper layers of the star.  When the shock reaches the 
surface of the star it abruptly accelerates, as shown in the velocity profiles of Figures 
\ref{fig:sbo1} and \ref{fig:sbo2}.  The shock also releases a brief, intense pulse of 
photons into the envelope.  This transient, which is mostly x-rays and hard UV, blows 
off the outer layers of the star, which detach from and accelerate ahead of the shock 
as we show at 8192 and 8407 s in the SMS1 velocities and at 8940 and 1.02e04 s in 
the SMS2 velocities.  This effect is more pronounced in SMS1 because it is easier for 
the radiation front to drive a precursor into the diffuse wind than the dense infall.  The 
advancing radiation front is visible as the flat plateau in gas temperature that extends 
from the outer edge of the shock into the surrounding medium.  The temperature to 
which the radiation heats the gas falls as the shock expands, cools, and its spectrum 
softens (note that the temperature of the shock itself is much higher).  

At breakout there are marked differences in the profiles of the two explosions as the 
shock plows into the envelope.  In both cases the shock accelerates but then slows 
down as it crashes out into the surrounding envelope, although the deceleration is
stronger in the dense infall.  But the SMS1 shock reaches much higher peak velocities
than the SMS2 shock.  This is partly due to the greater inertia of the infall and partly 
because the radiation front more easily blows off the outermost layers of the star in
the diffuse wind.  The radiation front also advances more quickly into the diffuse wind 
than the accretion flow.  On the other hand, when the shock breaks out into the dense 
envelope it heats it to much higher temperatures.  This hardens the spectrum of the 
shock and raises the temperature of the surrounding gas to higher temperatures than 
in SMS1, $\sim$ 100 eV instead of $\sim$ 50 eV.

\begin{figure}
\plotone{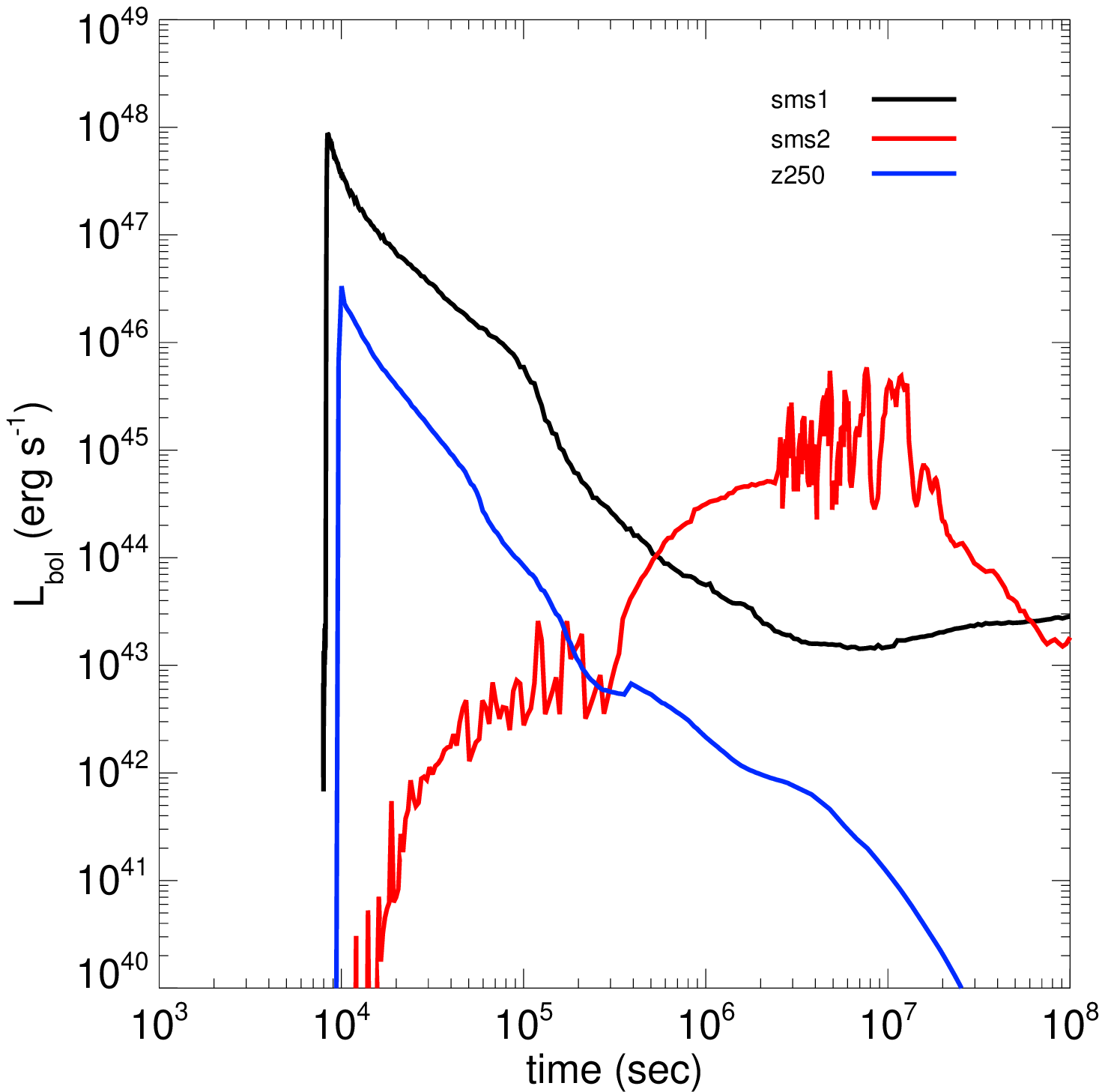}
\caption{Bolometric luminosities for SMS1 and SMS2, together with the light curve for 
the z250 PI SN from \citet{wet12b}.
\vspace{0.1in}}
\label{fig:lbol}
\end{figure}

\begin{figure}
\plotone{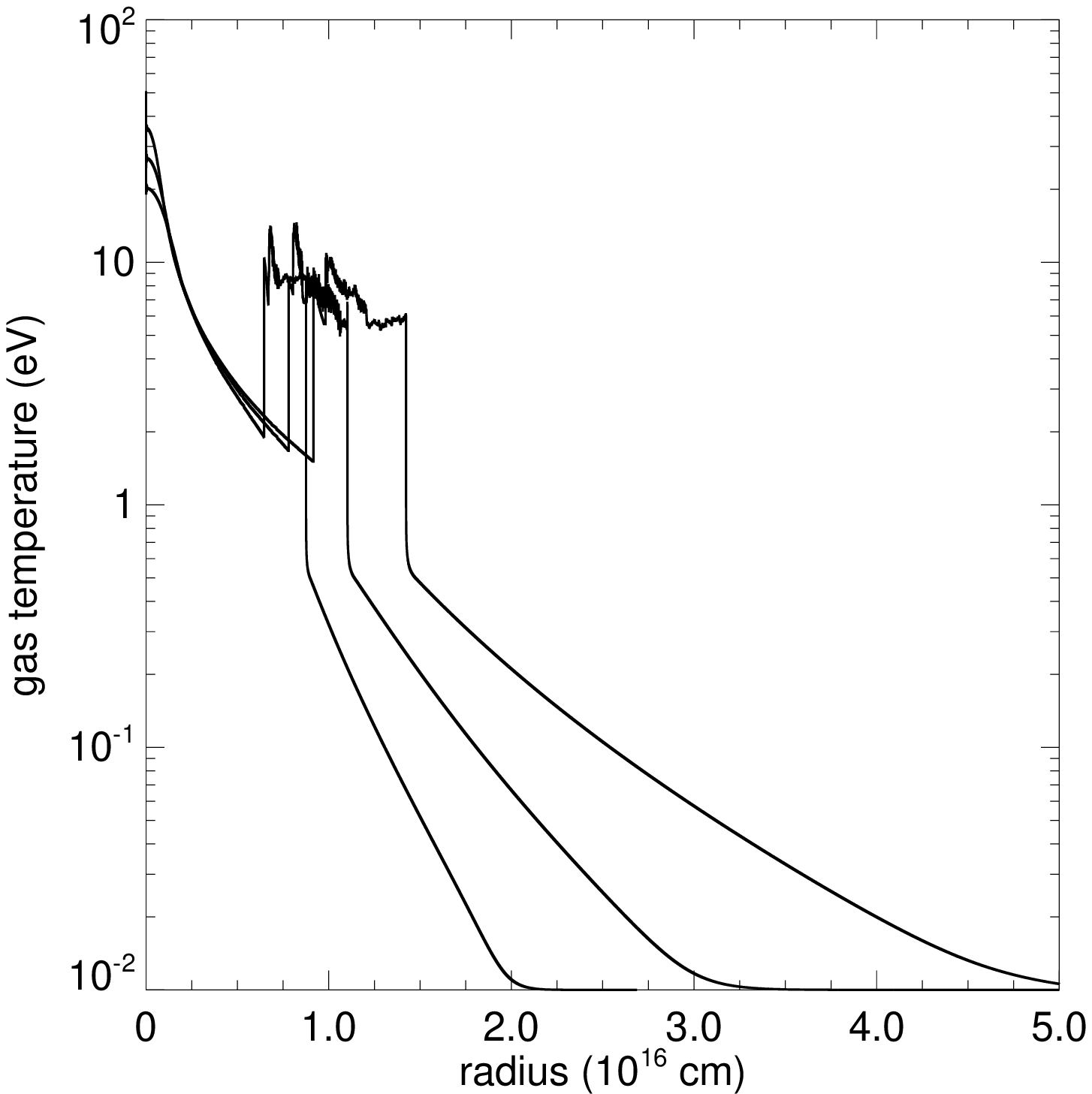}
\caption{Radiation breakout in SMS2:  gas temperatures at 5.78e6 s, 7.70e6 s, 
and 1.02e7 s.
\vspace{0.1in}}
\label{fig:rbo2}
\end{figure}

We show bolometric light curves for SMS1 and SMS2 in Figure \ref{fig:lbol}.  The 
light curve for SMS1 is similar to those of lower-mass Pop III PI SNe except that the 
explosion is approximately 40 times as luminous, as shown by the z250 light curve 
from \citet{wet12b}.  SMS1 exhibits the classic breakout transient, whose width is 
related to the light-crossing time of the star but is somewhat broader due to 
radiation-matter coupling effects as discussed in section 4.1 of \citet{wet12b}.  Its 
light curve is similar in structure to that of z250, which is also the explosion of a 
compact blue giant.  At early times the luminosity of SMS1 comes from the 
conversion of kinetic energy into thermal energy by the shock.  Later, ejecta cooling 
(not \Ni\ decay, since virtually none forms in these explosions) also contributes to its 
luminosity, in contrast to lower-mass PI SNe which are primarily powered by \Ni\ at 
later times.  As in the z-series PI SNe, there is a resurgence in luminosity at $\sim$ 
10$^7$ s that is again due to optical depth.  At this time the $\tau$ = 1 surface 
associated with the wavelength of peak emission in the spectrum has sunk to a hot 
layer deep in the ejecta, exposing it to the IGM and causing the SN to rebrighten.

\begin{figure*}
\plottwo{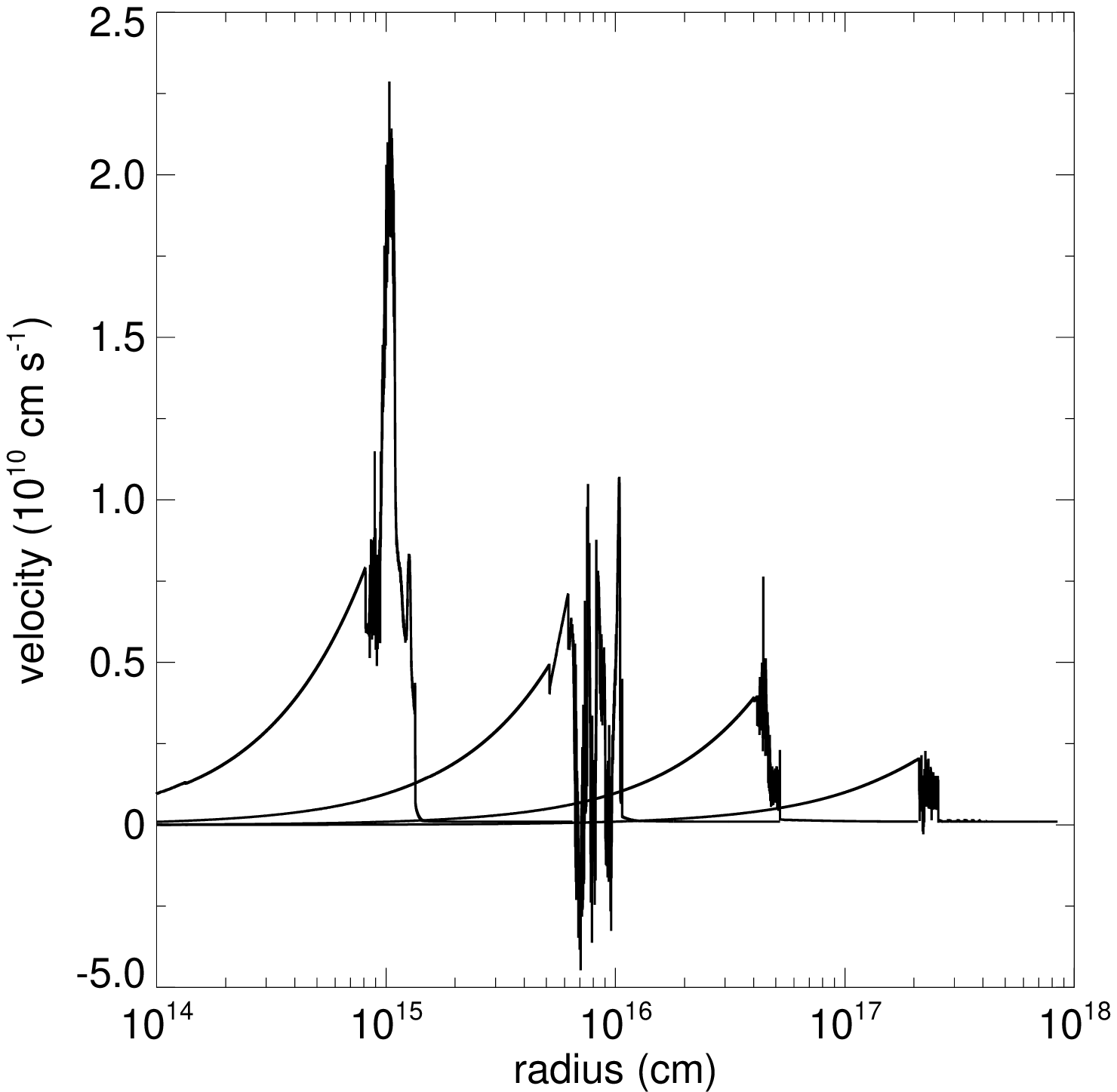}{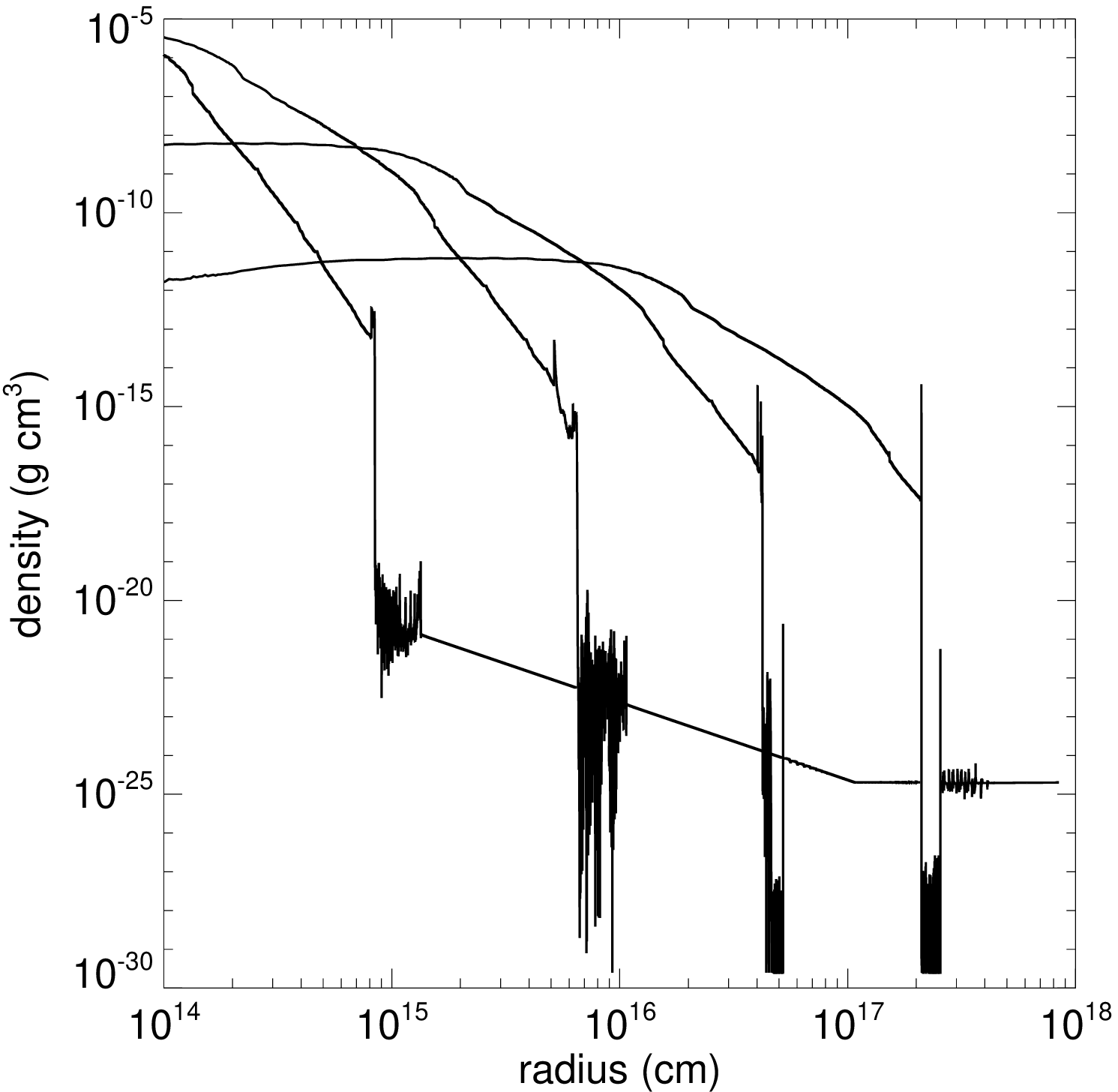}
\caption{Evolution of SMS1 at intermediate and late times.  Right:  velocities; from left to 
right:  10$^5$ s, 10$^6$ s, 10$^7$ s and 10$^8$ s.  Left:  densities at the same times.}
\label{fig:sms1}
\end{figure*}

\begin{figure*}
\plottwo{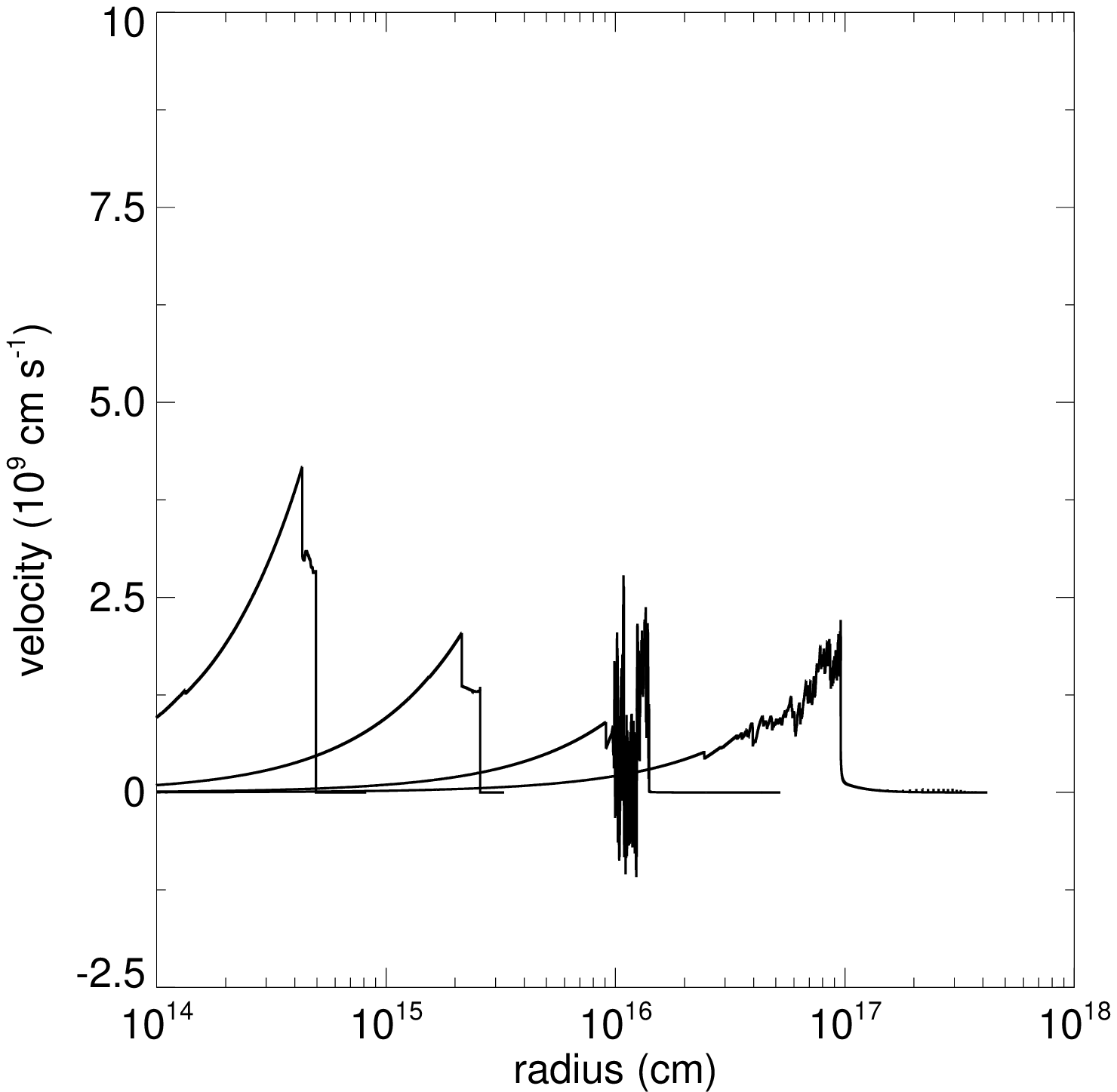}{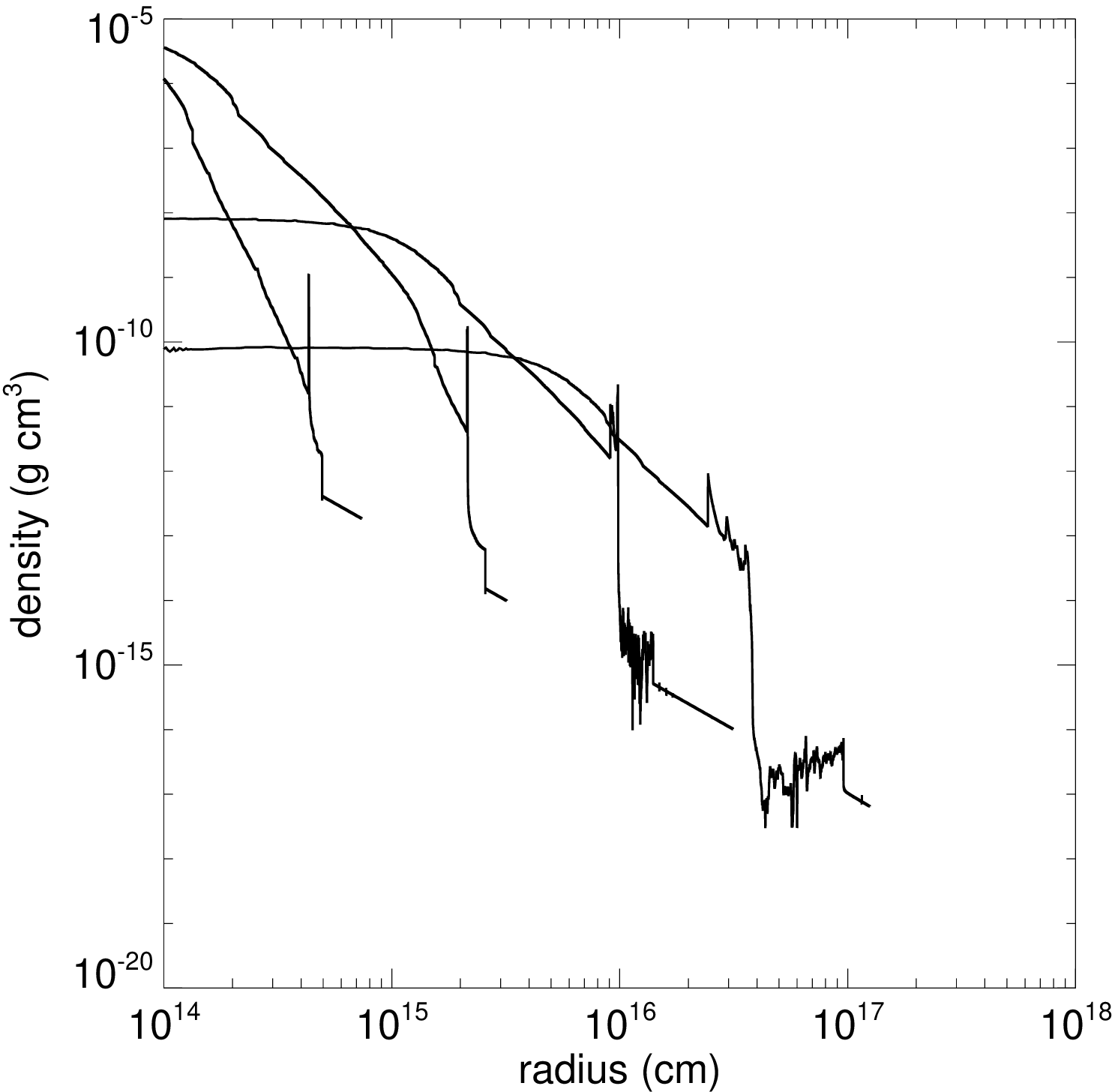}
\caption{Evolution of SMS2 at intermediate and late times.  Right:  velocities; from left to 
right:  10$^5$ s, 10$^6$ s, 10$^7$ s and 10$^8$ s.  Left:  densities at the same times.}
\label{fig:sms2}
\end{figure*}

Radiation breakout in SMS2 occurs far after shock breakout, at $\sim$ 3.0 $\times$ 10$
^6$ s as we show in Figure \ref{fig:lbol}.  Radiation escapes the dense envelope much 
later because of its large optical depth, and when it happens it is gradual, as we show 
in Figure \ref{fig:rbo2}.  Low-energy photons begin to leak out through the $\tau$ = 1 
surface for Thomson scattering at $\sim$ 2.6 $\times$ 10$^6$ s, and they are followed 
by more energetic photons by 5.6 $\times$ 10$^{6}$ s.  At this point the shock is much 
cooler because of the large amount of $PdV$ work it must perform on the dense shroud 
as it expands, but this results in an extremely luminous event in the NIR, as we discuss 
below.

After radiation breakout the shock appears to flicker until $\sim$ 1.7 $\times$ 10$^7$ s.  
This is due to radiative cooling and the cyclic formation and dissipation of a reverse shock 
in the ejecta.  As the shock plows up the envelope a reverse shock breaks free from the 
forward shock and is driven into the interior of the ejecta in the frame of the flow.  As the 
reverse shock detaches and recedes from the forward shock, it loses pressure support to 
radiative cooling by emission lines in the shocked gas and retreats back toward the 
forward shock. As the forward shock continues to sweep up the envelope a reverse shock 
again forms and backsteps from the forward shock.  The cyclic heating and cooling of 
shocked gas associated with the oscillation of the reverse shock, together with fluctuations
in opacities associated with these temperature cycles, cause the variations in luminosity 
from 2.5 $\times$ 10$^6$ - 1.7 $\times$ 10$^7$ s.  The period of oscillation is determined 
by cooling rates in the gas \citep{chev82,imam84,anet97} and is independent of the mass 
swept up by the ejecta.  Such ripples are also found in Lyman alpha emission by primordial 
SN remnants as they sweep up neutral gas in cosmological halos on larger scales 
\citep[note Figure 11 in][]{wet08a}.  The light curves of both SMS1 and SMS2 are easily 
distinguished from those of less massive Pop III PI SNe. 

We show velocity and density profiles for both explosions at intermediate to late times in
Figures \ref{fig:sms1} and \ref{fig:sms2}.  Multiple shocks are evident in the diffuse wind
just ahead of the shock in the SMS1 run at earlier times but they mostly dissipate by 3 yr, 
although some structures are still visible in the velocity.  These shocks are driven by the
propagation of radiation through the low-density wind ahead of the shock rather than by
the sweeping up of gas by the shock (indeed, the ejecta does not accumulate its own 
mass in ambient gas until it has grown to 6 pc).  The formation of a strong reverse shock 
due to plowed-up gas can be seen in the SMS2 velocity profiles from 10$^5$ s to 10$^6$ 
s.  By 10$^7$ s the surrounding wind has become sufficiently diffuse that the propagation
of radiation from the shock through it has created the same multiple shocks in it as in 
SMS1.  

\section{NIR Light Curves}

\begin{figure*}
\plottwo{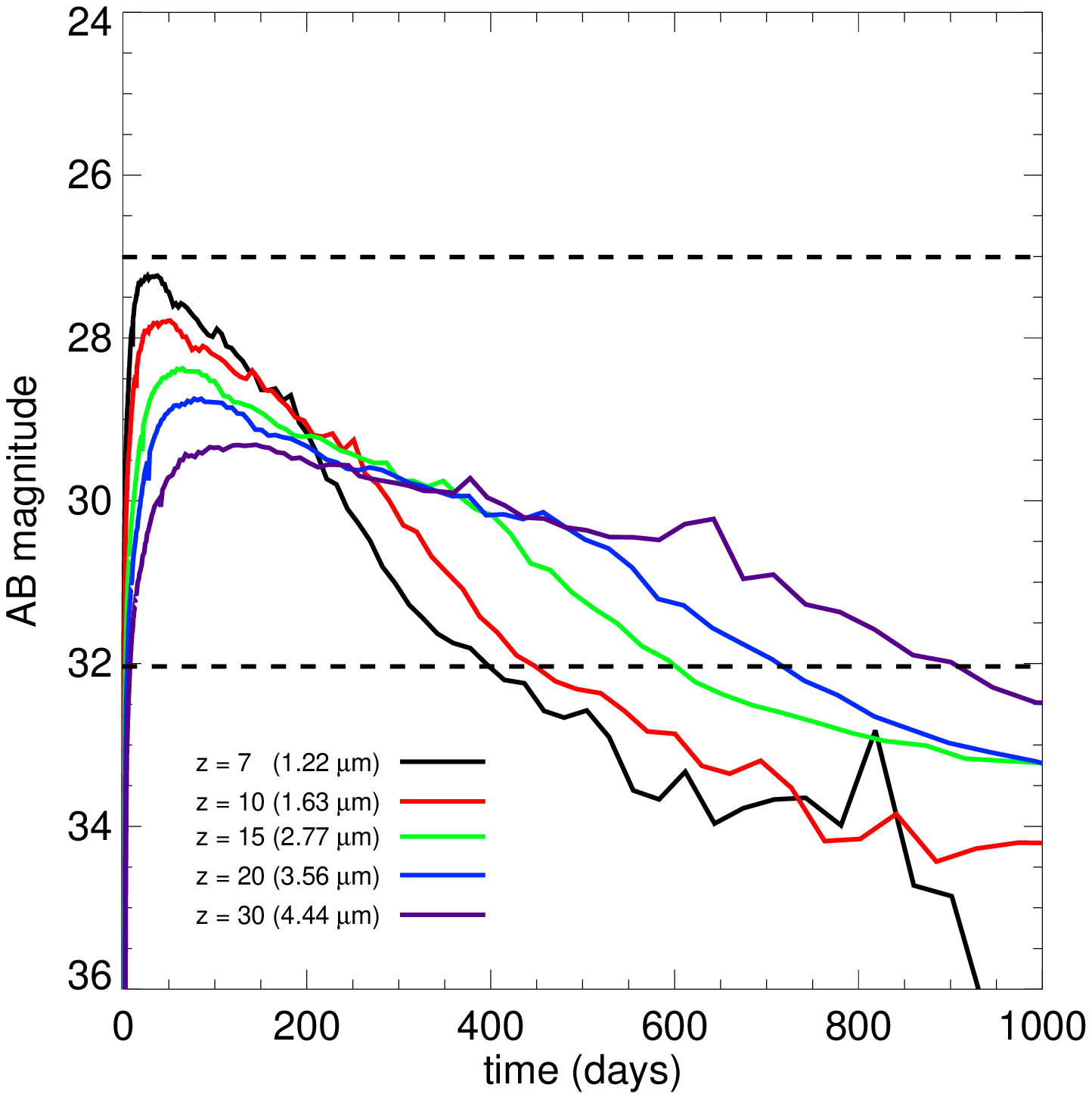}{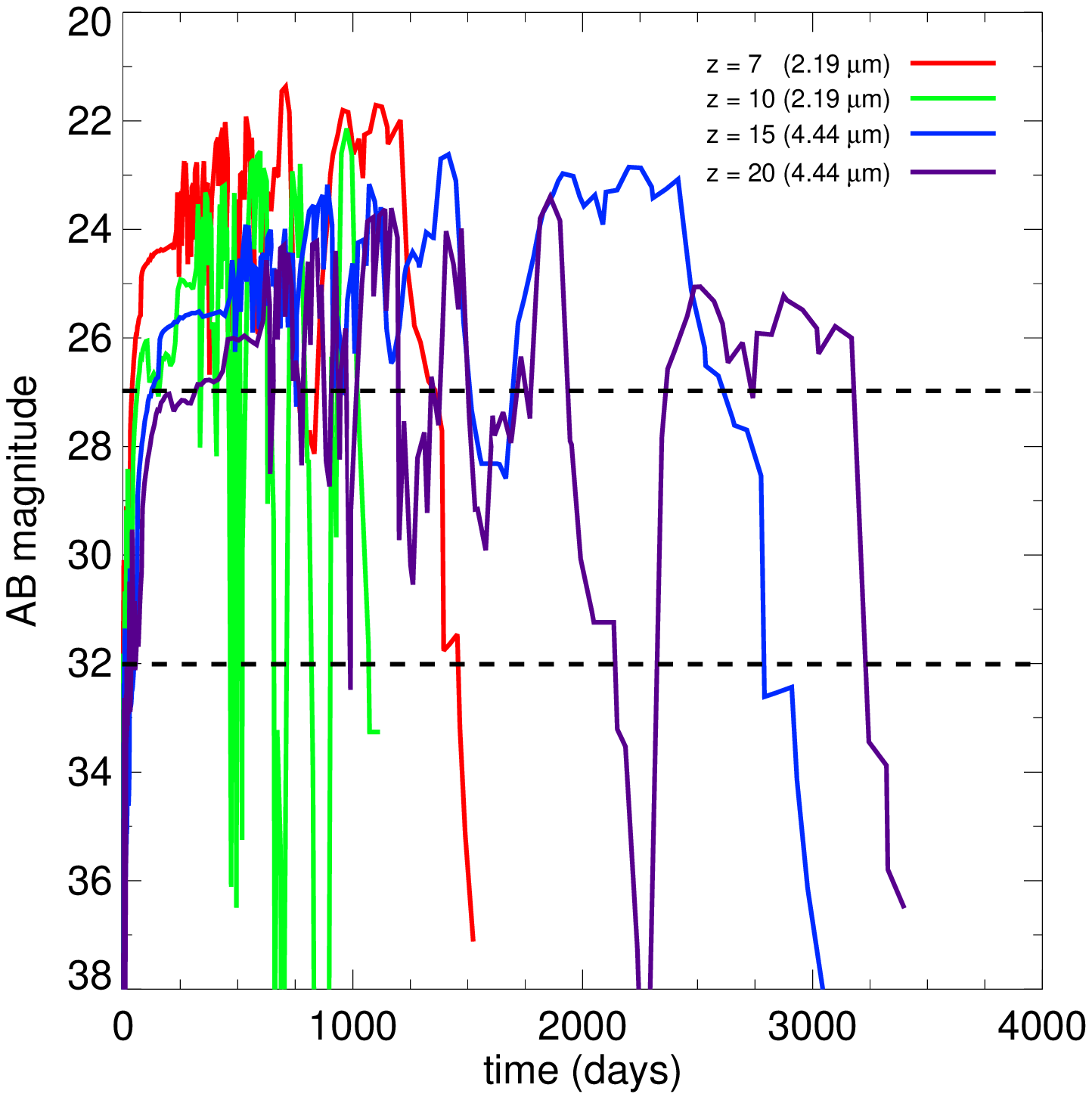}
\caption{\textit{JWST} NIRCam light curves for supermassive Pop III SNe in diffuse winds 
(SMS1, left panel) and dense envelopes (SMS2, right panel).  The horizontal dashed lines 
at AB magnitudes 32 and 27 are the photometry limits for \textit{JWST} and \textit{WFIRST}, 
respectively.}
\label{fig:NIRCam}
\end{figure*}

We calculate NIR light curves from our spectra with the photometry code developed by 
\citet{su11}.  Each spectrum is redshifted prior to removing the flux that is absorbed by 
intervening neutral hydrogen along the line of sight using the method of \citet{madau95}. 
The spectrum is then dimmed by the required cosmological factors for a specified 
redshift. The least sampled data is linearly interpolated between the input spectrum and 
filter curve to model the light curve in a given filter.  

\subsection{SMS1}

NIR luminosities are plotted for SMS1 at $z =$ 7, 10, 15, 20 and 30 in the left panel of 
Figure \ref{fig:NIRCam}. The SN will be visible to \textit{JWST} at all epochs for over 
1000 days but falls below the photometry limit of \textit{WFIRST} and \textit{WISH} at 
$z \gtrsim$ 7.  If spectrum stacking extends the detection limit of \textit{WFIRST} to 
AB magnitude 29 it could detect these explosions out to $z \sim$ 10.  SMS1 is quite 
luminous in the NIR, with peak magnitudes ranging from 27.5 at $z =$ 7 to 29.5 at $z 
=$ 20.  These light curves exhibit far more variability than their redshifted bolometric 
light curves might suggest, eliminating any possibility that these events would be 
mistaken for high-redshift protogalaxies.  This variation is due to the expansion and 
cooling of the fireball in the source frame.  

The NIR profiles of SMS1 are easily distinguished from those of the u-series and z-series 
PI SNe at all redshifts \citet[see Figures 10 and 11of][]{wet12b}. The SMS1 NIR light curves 
are similar in shape to those of z-series PI SNe, but the z-series luminosities are always 
several magnitudes dimmer at $z >$ 7.  The SMS1 light curves evolve with redshift as 
expected:  they broaden as $z$ increases and the optimum filter wavelength increases 
with redshift.  The NIR luminosities rise more quickly than they decline so these events are 
most easily detected in their early stages, but they nonetheless exhibit enough variability at
any stage to be found in multi-year baseline searches.

\subsection{SMS2}

We show NIR luminosities for SMS2 at $z =$ 7, 10, 15, 20 and 30 in the right panel of 
Figure \ref{fig:NIRCam}.  They are quite different from those of SMS1.  Consistent with 
radiation breakout from the shroud at $\sim$ 20 days, no NIR signal is observed from 
these events until 100 - 150 days at $z >$ 7.  This explosion eventually becomes 
hundreds of times brighter in the NIR than SMS1, with peak AB magnitudes from 21 at 
$z =$ 7 to 23 at $z =$ 15. It is visible to \textit{JWST} for 1000 - 3000 days out to $z 
\sim$ 20 and to \textit{WFIRST} and \textit{WISH} for 1000 days out to $z \sim $ 15 - 20.  
We also note that \textit{Euclid}, with a photometry limit of AB mag 24 at 2.2 $\mu$m, 
can detect SMS2 for $\sim$ 1000 days at $z =$ 10 - 15, the likely epoch of these events.  
The much higher NIR luminosities are due to the large radius of the fireball at radiation 
breakout and the relatively low temperature of the shock at this radius ($\sim$ 10 eV) 
because of the $PdV$ work the fireball must do against the dense envelope.  These 
lower temperatures drive the redshifted peak of the shock's spectrum into the NIR in the 
observer frame.  The relative magnitudes of the three light curves are properly ordered 
in redshift.  The ripples in luminosity have much shorter periods than those in the 
bolometric luminosity in Figure \ref{fig:lbol} and are likely due to opacity fluctuations in 
the shock.    

In sum, SMS explosions in both diffuse winds and dense envelopes will be visible in 
\textit{JWST} NIR deep fields out to $z \gtrsim$ 30 but only the latter will be visible to 
all-sky NIR surveys by \textit{Euclid}, \textit{WFIRST}, and \textit{WISH}.  But they will 
be visible at $z \sim$ 15, which is when they likely begin to occur.  The fact that these 
NIR profiles change considerably with circumstellar envelope suggests that they will 
be powerful probes of the environments of such explosions.  It is worth noting that 
even fully shrouded explosions will be visible at the earliest epochs.  The envelopes 
we have chosen should bracket those in which SMS PI SNe will occur, so the NIR 
signals of actual explosions may be intermediate to those of these two events.  Given
the massive infall rates required to form supermassive Pop III stars, it is unlikely these
stars fully disperse their accretion envelopes in their lifetimes \citep{jlj12a}, and so we
expect their SN light curves to be closer to SMS2 than SMS1 in brightness.  As noted 
earlier, both SMS1 and SMS2 are easily distinguished from low-mass Pop III PI SNe 
as well as core-collapse SNe \citep{wet12c} and Type IIn SNe \citep{moriya10,wet12e}.

\section{Conclusion}

The discovery of supermassive Pop III PI SNe would confirm for the first time that massive 
fragments capable of collapsing to 10$^4$ -- 10$^5$ \Ms\ SMBH seeds do form in primeval
galaxies at high redshift.  Although the rate of such events remains unknown, it might be 
thought that they are very rare because supermassive fragments must fall into a relatively 
narrow mass range to actually become stars and because few protogalaxies form in LW 
backgrounds capable of fully sterilizing them of H$_2$.  However, recent developments 
suggest that these processes were more frequent than previously thought.  

First, new simulations indicate that the assembly of protogalaxies in strong LW  backgrounds 
may have been relatively common, yielding higher rates of SMBH seed production than might 
naively be inferred from the number density of $z\sim$ 7 quasars, $\sim$ 1 Gpc$^{-3}$ \citep{
agarw12}. The sustained exponential growth required to reach such masses depended on the 
topology of cold flows over cosmic time \citep{dm12}, so the scarcity of such flows may have 
governed the density of $z \sim$ 7 quasars, not the rate of seed formation.  Second, rotation 
could broaden the mass range over which supermassive stars encounter the pair instability
by enhancing mixing and more rapidly building up massive He cores \citep{cw12}.  Greater 
mass ranges imply larger event rates.  

A reasonable upper limit to SMS PI SN event rates are those of 140 - 260 \Ms\ Pop III PI SNe, 
which \citet{wet12b} and others find to be $\sim$ 10$^{-2}$ yr$^{-1}$ deg$^{-2}$ at $z \gtrsim$ 
10, which implies all-sky rates of up to $\sim$ 10$^3$ yr$^{-1}$ \citep{wa05,wl05,oet05,tfs07,
wet08b,tss09,wet10,get10,maio11,hum12,jdk12,wise12}.  Although actual SMS PI SN event 
rates may be well below this limit, precluding their detection by \textit{JWST}, they will clearly 
be bright enough to appear in wide field campaigns.  They might also be detected in present 
NIR all-sky surveys by the \textit{Subaru Hyper Suprime-Cam} at $z \gtrsim$ 3 but further 
calculations will be necessary to confirm this \citep[e.g.,][]{tomin11,tet12,moriya12}. Synchrotron 
emission from Pop III SNe at $z \gtrsim$ 10 can be detected at 21 cm by existing observatories 
such as \textit{eVLA} and \textit{eMerlin} and future ones such as the \textit{Square Kilometer 
Array} (\textit{SKA}) \citep{mw12}.  SMS PI SN explosions in dense envelopes may likewise 
generate enough synchrotron emission to be discovered in radio surveys.

We note that the formation of somewhat less massive SMS stars ($\sim 10^3 - 10^4$ \Ms) may
lead to highly energetic ($E_{\gamma, \mathrm{iso}} \sim$ 10$^{55}$ - 10$^{57}$ erg s$^{-1}$) 
gamma-ray bursts \citep[GRBs;][]{suwa11,nsi12} \citep[see also][]{wet08c,met12a}.  They are 
characterized by extended prompt emission because of the large reservoirs that can drive the 
central engine and that are necessary for the GRB jet to puncture the large envelopes of such
stars.  The afterglows of very massive Pop III GRBs can appear in both current and future radio 
surveys by eVLA, eMERLIN and the Square Kilometer Array \citep[SKA;][]{ds11}.  They will also 
be visible in future all-sky NIR campaigns by WFIRST and WISH.

Given that most supermassive stars will still directly collapse to BH, could there be other ways 
of detecting SMBH seed formation in protogalaxies?  Past studies have shown that collapsing 
supermassive stars become extremely luminous in thermal neutrino emission as the central BH 
forms, with energies of $\sim$ 10 MeV \citep{fuller86,sfh98,montero12}.  The prospects for 
detecting such neutrinos depends on the initial mass and entropy of the core. Although the total 
energy emitted from these massive stars increases linearly with mass, the entropy of their cores 
also increases with mass.  Higher entropies lead to larger proto-black holes with lower peak 
densities and lower temperatures. \citet{fh11} found that the neutrino luminosity does not increase 
much with mass for stars above 10,000 \Ms.  The mean electron neutrino energy for stars above 
10,000 \Ms\ is below 6 MeV and the $\mu_{\tau}$ energy is not much higher.  The collapse of 
such cores would be difficult to observe with neutrino detectors.  However, if the core entropy is 
closer to that of a 1,000 \Ms\ star, the luminosity peaks more dramatically.  Although even for 
these cores the mean electron neutrino energy is $\sim 7-8$ MeV, the $\mu_{\tau}$ energy lies 
in the 20-30 MeV range and would be more easily detected after cosmological redshifting.  More 
detailed calculations are needed to be certain, but these cores would likely contribute to the 
neutrino background in detectors such as IceCube.  If the density profile of the collapsing star 
also imposes a unique spectrum on the neutrino flux, it would facilitate its extraction from noise.  

As noted in the Introduction, collapse may also lead to the formation of a black hole accretion 
disk system without the creation of a star, with nuclear burning near the event horizon whose 
products could be expelled out into the halo by a jet \citep[e.g.,][]{smh06,sm08}.  The 
nucleosynthetic signature of this process, which could be imprinted on stars that later form in 
the protogalaxy, depends on the temperature of the disk and hence the radius of the BH.  It 
can therefore provide a diagnostic of the mass of the SMBH seed at birth, since massive BH 
with large event horizons burn at lower temperatures and yield chemical abundances that are 
distinct from those of smaller disks, which can burn all the way to Ni.  Ancient, dim metal-poor 
stars bearing the ashes of this process could soon be discovered in ongoing surveys in the 
Galactic halo \citep[e.g.,][]{Cayrel2004,bc05,fet05,Lai2008,mbh03,ss07,bsmith09,chiaki12,
ritt12}.  The collapse of a supermassive star could also emit gravity waves (GWs) that might be 
detected in existing or future GW facilities \citep[e.g.,][]{fhh02,fn11}.  These multi-messenger 
events, together with the most energetic supernovae in the universe, may soon herald the 
births of the first quasars.

\acknowledgments

We thank the anonymous referee, whose comments improved the quality of this paper.  DJW 
thanks George Fuller, John J. Cherry and Jarrett Johnson for enlightening discussions on the 
evolution of supermassive stars and Terrance Strother for running some of the calculations.  
He acknowledges support from the Bruce and Astrid McWilliams Center for Cosmology at 
Carnegie Mellon University and from the Baden-W\"{u}rttemberg-Stiftung by contract research 
via the programme Internationale Spitzenforschung II (grant P- LS-SPII/18).  AH and KC were 
supported by the US Department of Energy under contracts DE-FC02-01ER41176, 
FC02-09ER41618 (SciDAC), and DE-FG02-87ER40328.  MS thanks Marcia Rieke for making 
available the NIRCam filter curves and was partially supported by NASA JWST grant 
NAG5-12458.  Work at LANL was done under the auspices of the National Nuclear Security 
Administration of the U.S. Department of Energy at Los Alamos National Laboratory under 
Contract No. DE-AC52-06NA25396.  All RAGE and SPECTRUM calculations were performed 
on Institutional Computing (IC) and Yellow network platforms at LANL (Mustang, Pinto, Conejo, 
Lobo and Yellowrail).

\bibliographystyle{apj}
\bibliography{refs}

\begin{thebibliography}{158}
\expandafter\ifx\csname natexlab\endcsname\relax\def\natexlab#1{#1}\fi

\bibitem[{{Abel} {et~al.}(2000){Abel}, {Bryan}, \& {Norman}}]{abn00}
{Abel}, T., {Bryan}, G.~L., \& {Norman}, M.~L. 2000, \apj, 540, 39

\bibitem[{{Abel} {et~al.}(2002){Abel}, {Bryan}, \& {Norman}}]{abn02}
---. 2002, Science, 295, 93

\bibitem[{{Abel} {et~al.}(2007){Abel}, {Wise}, \& {Bryan}}]{awb07}
{Abel}, T., {Wise}, J.~H., \& {Bryan}, G.~L. 2007, \apjl, 659, L87

\bibitem[{{Agarwal} {et~al.}(2012){Agarwal}, {Khochfar}, {Johnson}, {Neistein},
  {Dalla Vecchia}, \& {Livio}}]{agarw12}
{Agarwal}, B., {Khochfar}, S., {Johnson}, J.~L., {Neistein}, E., {Dalla
  Vecchia}, C., \& {Livio}, M. 2012, \mnras, 425, 2854

\bibitem[{{Almgren} {et~al.}(2010){Almgren}, {Beckner}, {Bell}, {Day},
  {Howell}, {Joggerst}, {Lijewski}, {Nonaka}, {Singer}, \&
  {Zingale}}]{Almgren2010}
{Almgren}, A.~S., {Beckner}, V.~E., {Bell}, J.~B., {Day}, M.~S., {Howell},
  L.~H., {Joggerst}, C.~C., {Lijewski}, M.~J., {Nonaka}, A., {Singer}, M., \&
  {Zingale}, M. 2010, \apj, 715, 1221

\bibitem[{{Alvarez} {et~al.}(2006){Alvarez}, {Bromm}, \& {Shapiro}}]{abs06}
{Alvarez}, M.~A., {Bromm}, V., \& {Shapiro}, P.~R. 2006, \apj, 639, 621

\bibitem[{{Alvarez} {et~al.}(2009){Alvarez}, {Wise}, \& {Abel}}]{awa09}
{Alvarez}, M.~A., {Wise}, J.~H., \& {Abel}, T. 2009, \apjl, 701, L133

\bibitem[{{Anninos} {et~al.}(1997){Anninos}, {Zhang}, {Abel}, \&
  {Norman}}]{anet97}
{Anninos}, P., {Zhang}, Y., {Abel}, T., \& {Norman}, M.~L. 1997, New Astronomy,
  2, 209

\bibitem[{{Baraffe} {et~al.}(2001){Baraffe}, {Heger}, \& {Woosley}}]{Baraffe01}
{Baraffe}, I., {Heger}, A., \& {Woosley}, S.~E. 2001, \apj, 550, 890

\bibitem[{{Barkat} {et~al.}(1967){Barkat}, {Rakavy}, \& {Sack}}]{brk67}
{Barkat}, Z., {Rakavy}, G., \& {Sack}, N. 1967, Physical Review Letters, 18,
  379

\bibitem[{{Beers} \& {Christlieb}(2005)}]{bc05}
{Beers}, T.~C. \& {Christlieb}, N. 2005, \araa, 43, 531

\bibitem[{{Begelman}(2010)}]{begel10}
{Begelman}, M.~C. 2010, \mnras, 402, 673

\bibitem[{{Begelman} {et~al.}(2008){Begelman}, {Rossi}, \&
  {Armitage}}]{begel08}
{Begelman}, M.~C., {Rossi}, E.~M., \& {Armitage}, P.~J. 2008, \mnras, 387, 1649

\bibitem[{{Begelman} {et~al.}(2006){Begelman}, {Volonteri}, \&
  {Rees}}]{begel06}
{Begelman}, M.~C., {Volonteri}, M., \& {Rees}, M.~J. 2006, \mnras, 370, 289

\bibitem[{{Bromm} {et~al.}(1999){Bromm}, {Coppi}, \& {Larson}}]{bcl99}
{Bromm}, V., {Coppi}, P.~S., \& {Larson}, R.~B. 1999, \apjl, 527, L5

\bibitem[{{Bromm} {et~al.}(2002){Bromm}, {Coppi}, \& {Larson}}]{bcl02}
---. 2002, \apj, 564, 23

\bibitem[{{Bromm} \& {Loeb}(2003)}]{bl03}
{Bromm}, V. \& {Loeb}, A. 2003, \apj, 596, 34

\bibitem[{{Bromm} {et~al.}(2003){Bromm}, {Yoshida}, \& {Hernquist}}]{byh03}
{Bromm}, V., {Yoshida}, N., \& {Hernquist}, L. 2003, \apjl, 596, L135

\bibitem[{{Cayrel} {et~al.}(2004){Cayrel}, {Depagne}, {Spite}, {Hill}, {Spite},
  {Fran{\c c}ois}, {Plez}, {Beers}, {Primas}, {Andersen}, {Barbuy},
  {Bonifacio}, {Molaro}, \& {Nordstr{\"o}m}}]{Cayrel2004}
{Cayrel}, R., {Depagne}, E., {Spite}, M., {Hill}, V., {Spite}, F., {Fran{\c
  c}ois}, P., {Plez}, B., {Beers}, T., {Primas}, F., {Andersen}, J., {Barbuy},
  B., {Bonifacio}, P., {Molaro}, P., \& {Nordstr{\"o}m}, B. 2004, \aap, 416,
  1117

\bibitem[{{Chatzopoulos} \& {Wheeler}(2012)}]{cw12}
{Chatzopoulos}, E. \& {Wheeler}, J.~C. 2012, \apj, 748, 42

\bibitem[{{Chatzopoulos} {et~al.}(2013){Chatzopoulos}, {Wheeler}, \&
  {Couch}}]{cwc13}
{Chatzopoulos}, E., {Wheeler}, J.~C., \& {Couch}, S.~M. 2013, arXiv:1308.4660

\bibitem[{{Chen} {et~al.}(2011){Chen}, {Heger}, \& {Almgren}}]{chen11}
{Chen}, K.-J., {Heger}, A., \& {Almgren}, A.~S. 2011, Computer Physics
  Communications, 182, 254

\bibitem[{{Chevalier} \& {Imamura}(1982)}]{chev82}
{Chevalier}, R.~A. \& {Imamura}, J.~N. 1982, \apj, 261, 543

\bibitem[{{Chiaki} {et~al.}(2013){Chiaki}, {Yoshida}, \& {Kitayama}}]{chiaki12}
{Chiaki}, G., {Yoshida}, N., \& {Kitayama}, T. 2013, \apj, 762, 50

\bibitem[{{Choi} {et~al.}(2013){Choi}, {Shlosman}, \& {Begelman}}]{choi13}
{Choi}, J.-H., {Shlosman}, I., \& {Begelman}, M.~C. 2013, \apj, 774, 149

\bibitem[{{Clark} {et~al.}(2011){Clark}, {Glover}, {Smith}, {Greif}, {Klessen},
  \& {Bromm}}]{clark11}
{Clark}, P.~C., {Glover}, S.~C.~O., {Smith}, R.~J., {Greif}, T.~H., {Klessen},
  R.~S., \& {Bromm}, V. 2011, Science, 331, 1040

\bibitem[{{de Souza} {et~al.}(2013){de Souza}, {Ishida}, {Johnson}, {Whalen},
  \& {Mesinger}}]{ds13}
{de Souza}, R.~S., {Ishida}, E.~E.~O., {Johnson}, J.~L., {Whalen}, D.~J., \&
  {Mesinger}, A. 2013, arXiv:1306.4984

\bibitem[{{de Souza} {et~al.}(2011{\natexlab{a}}){de Souza}, {Rodrigues},
  {Ishida}, \& {Opher}}]{ds11a}
{de Souza}, R.~S., {Rodrigues}, L.~F.~S., {Ishida}, E.~E.~O., \& {Opher}, R.
  2011{\natexlab{a}}, \mnras, 415, 2969

\bibitem[{{de Souza} {et~al.}(2011{\natexlab{b}}){de Souza}, {Yoshida}, \&
  {Ioka}}]{ds11}
{de Souza}, R.~S., {Yoshida}, N., \& {Ioka}, K. 2011{\natexlab{b}}, \aap, 533,
  A32

\bibitem[{{Di Matteo} {et~al.}(2012){Di Matteo}, {Khandai}, {DeGraf}, {Feng},
  {Croft}, {Lopez}, \& {Springel}}]{dm12}
{Di Matteo}, T., {Khandai}, N., {DeGraf}, C., {Feng}, Y., {Croft}, R.~A.~C.,
  {Lopez}, J., \& {Springel}, V. 2012, \apjl, 745, L29

\bibitem[{{Djorgovski} {et~al.}(2008){Djorgovski}, {Volonteri}, {Springel},
  {Bromm}, \& {Meylan}}]{brmvol08}
{Djorgovski}, S.~G., {Volonteri}, M., {Springel}, V., {Bromm}, V., \& {Meylan},
  G. 2008, in The Eleventh Marcel Grossmann Meeting On Recent Developments in
  Theoretical and Experimental General Relativity, Gravitation and Relativistic
  Field Theories, ed. {H.~Kleinert, R.~T.~Jantzen, \& R.~Ruffini}, 340--367

\bibitem[{{Frebel} {et~al.}(2005){Frebel}, {Aoki}, {Christlieb}, {Ando},
  {Asplund}, {Barklem}, {Beers}, {Eriksson}, {Fechner}, {Fujimoto}, {Honda},
  {Kajino}, {Minezaki}, {Nomoto}, {Norris}, {Ryan}, {Takada-Hidai},
  {Tsangarides}, \& {Yoshii}}]{fet05}
{Frebel}, A., {Aoki}, W., {Christlieb}, N., {Ando}, H., {Asplund}, M.,
  {Barklem}, P.~S., {Beers}, T.~C., {Eriksson}, K., {Fechner}, C., {Fujimoto},
  M.~Y., {Honda}, S., {Kajino}, T., {Minezaki}, T., {Nomoto}, K., {Norris},
  J.~E., {Ryan}, S.~G., {Takada-Hidai}, M., {Tsangarides}, S., \& {Yoshii}, Y.
  2005, \nat, 434, 871

\bibitem[{{Frey} {et~al.}(2013){Frey}, {Even}, {Whalen}, {Fryer}, {Hungerford},
  {Fontes}, \& {Colgan}}]{fet12}
{Frey}, L.~H., {Even}, W., {Whalen}, D.~J., {Fryer}, C.~L., {Hungerford},
  A.~L., {Fontes}, C.~J., \& {Colgan}, J. 2013, \apjs, 204, 16

\bibitem[{{Fryer} \& {Heger}(2011)}]{fh11}
{Fryer}, C.~L. \& {Heger}, A. 2011, Astronomische Nachrichten, 332, 408

\bibitem[{{Fryer} {et~al.}(2002){Fryer}, {Holz}, \& {Hughes}}]{fhh02}
{Fryer}, C.~L., {Holz}, D.~E., \& {Hughes}, S.~A. 2002, \apj, 565, 430

\bibitem[{{Fryer} \& {New}(2011)}]{fn11}
{Fryer}, C.~L. \& {New}, K.~C.~B. 2011, Living Reviews in Relativity, 14, 1

\bibitem[{{Fryer} {et~al.}(2010){Fryer}, {Whalen}, \& {Frey}}]{fwf10}
{Fryer}, C.~L., {Whalen}, D.~J., \& {Frey}, L. 2010, in American Institute of
  Physics Conference Series, Vol. 1294, American Institute of Physics
  Conference Series, ed. D.~J. {Whalen}, V.~{Bromm}, \& N.~{Yoshida}, 70--75

\bibitem[{{Fuller} {et~al.}(1986){Fuller}, {Woosley}, \& {Weaver}}]{fuller86}
{Fuller}, G.~M., {Woosley}, S.~E., \& {Weaver}, T.~A. 1986, \apj, 307, 675

\bibitem[{{Gal-Yam} {et~al.}(2009){Gal-Yam}, {Mazzali}, {Ofek}, {Nugent},
  {Kulkarni}, {Kasliwal}, {Quimby}, {Filippenko}, {Cenko}, {Chornock},
  {Waldman}, {Kasen}, {Sullivan}, {Beshore}, {Drake}, {Thomas}, {Bloom},
  {Poznanski}, {Miller}, {Foley}, {Silverman}, {Arcavi}, {Ellis}, \&
  {Deng}}]{gy09}
{Gal-Yam}, A., {Mazzali}, P., {Ofek}, E.~O., {Nugent}, P.~E., {Kulkarni},
  S.~R., {Kasliwal}, M.~M., {Quimby}, R.~M., {Filippenko}, A.~V., {Cenko},
  S.~B., {Chornock}, R., {Waldman}, R., {Kasen}, D., {Sullivan}, M., {Beshore},
  E.~C., {Drake}, A.~J., {Thomas}, R.~C., {Bloom}, J.~S., {Poznanski}, D.,
  {Miller}, A.~A., {Foley}, R.~J., {Silverman}, J.~M., {Arcavi}, I., {Ellis},
  R.~S., \& {Deng}, J. 2009, \nat, 462, 624

\bibitem[{{Gardner} {et~al.}(2006){Gardner}, {Mather}, {Clampin}, {Doyon},
  {Greenhouse}, {Hammel}, {Hutchings}, {Jakobsen}, {Lilly}, {Long}, {Lunine},
  {McCaughrean}, {Mountain}, {Nella}, {Rieke}, {Rieke}, {Rix}, {Smith},
  {Sonneborn}, {Stiavelli}, {Stockman}, {Windhorst}, \& {Wright}}]{jwst06}
{Gardner}, J.~P., {Mather}, J.~C., {Clampin}, M., {Doyon}, R., {Greenhouse},
  M.~A., {Hammel}, H.~B., {Hutchings}, J.~B., {Jakobsen}, P., {Lilly}, S.~J.,
  {Long}, K.~S., {Lunine}, J.~I., {McCaughrean}, M.~J., {Mountain}, M.,
  {Nella}, J., {Rieke}, G.~H., {Rieke}, M.~J., {Rix}, H.-W., {Smith}, E.~P.,
  {Sonneborn}, G., {Stiavelli}, M., {Stockman}, H.~S., {Windhorst}, R.~A., \&
  {Wright}, G.~S. 2006, \ssr, 123, 485

\bibitem[{{Gittings} {et~al.}(2008){Gittings}, {Weaver}, {Clover}, {Betlach},
  {Byrne}, {Coker}, {Dendy}, {Hueckstaedt}, {New}, {Oakes}, {Ranta}, \&
  {Stefan}}]{rage}
{Gittings}, M., {Weaver}, R., {Clover}, M., {Betlach}, T., {Byrne}, N.,
  {Coker}, R., {Dendy}, E., {Hueckstaedt}, R., {New}, K., {Oakes}, W.~R.,
  {Ranta}, D., \& {Stefan}, R. 2008, Computational Science and Discovery, 1,
  015005

\bibitem[{{Glover}(2013)}]{glov12}
{Glover}, S. 2013, in Astrophysics and Space Science Library, Vol. 396,
  Astrophysics and Space Science Library, ed. T.~{Wiklind}, B.~{Mobasher}, \&
  V.~{Bromm}, 103

\bibitem[{{Greif} {et~al.}(2012){Greif}, {Bromm}, {Clark}, {Glover}, {Smith},
  {Klessen}, {Yoshida}, \& {Springel}}]{get12}
{Greif}, T.~H., {Bromm}, V., {Clark}, P.~C., {Glover}, S.~C.~O., {Smith},
  R.~J., {Klessen}, R.~S., {Yoshida}, N., \& {Springel}, V. 2012, \mnras, 424,
  399

\bibitem[{{Greif} {et~al.}(2010){Greif}, {Glover}, {Bromm}, \&
  {Klessen}}]{get10}
{Greif}, T.~H., {Glover}, S.~C.~O., {Bromm}, V., \& {Klessen}, R.~S. 2010,
  \apj, 716, 510

\bibitem[{{Greif} {et~al.}(2008){Greif}, {Johnson}, {Klessen}, \&
  {Bromm}}]{get08}
{Greif}, T.~H., {Johnson}, J.~L., {Klessen}, R.~S., \& {Bromm}, V. 2008,
  \mnras, 387, 1021

\bibitem[{{Greif} {et~al.}(2011){Greif}, {Springel}, {White}, {Glover},
  {Clark}, {Smith}, {Klessen}, \& {Bromm}}]{get11}
{Greif}, T.~H., {Springel}, V., {White}, S.~D.~M., {Glover}, S.~C.~O., {Clark},
  P.~C., {Smith}, R.~J., {Klessen}, R.~S., \& {Bromm}, V. 2011, \apj, 737, 75

\bibitem[{{Heger} \& {Chen}(2013)}]{heg13}
{Heger}, A. \& {Chen}, K.-J. 2013, \apj, in prep

\bibitem[{{Heger} \& {Woosley}(2002)}]{hw02}
{Heger}, A. \& {Woosley}, S.~E. 2002, \apj, 567, 532

\bibitem[{{Heger} \& {Woosley}(2010)}]{hw10}
---. 2010, \apj, 724, 341

\bibitem[{{Hirano} {et~al.}(2013){Hirano}, {Hosokawa}, {Yoshida}, {Umeda},
  {Omukai}, {Chiaki}, \& {Yorke}}]{hir13}
{Hirano}, S., {Hosokawa}, T., {Yoshida}, N., {Umeda}, H., {Omukai}, K.,
  {Chiaki}, G., \& {Yorke}, H.~W. 2013, arXiv:1308.4456

\bibitem[{{Hosokawa} {et~al.}(2011){Hosokawa}, {Omukai}, {Yoshida}, \&
  {Yorke}}]{hos11}
{Hosokawa}, T., {Omukai}, K., {Yoshida}, N., \& {Yorke}, H.~W. 2011, Science,
  334, 1250

\bibitem[{{Hosokawa} {et~al.}(2012){Hosokawa}, {Yoshida}, {Omukai}, \&
  {Yorke}}]{hos12}
{Hosokawa}, T., {Yoshida}, N., {Omukai}, K., \& {Yorke}, H.~W. 2012, \apjl,
  760, L37

\bibitem[{{Hummel} {et~al.}(2012){Hummel}, {Pawlik}, {Milosavljevi{\'c}}, \&
  {Bromm}}]{hum12}
{Hummel}, J.~A., {Pawlik}, A.~H., {Milosavljevi{\'c}}, M., \& {Bromm}, V. 2012,
  \apj, 755, 72

\bibitem[{{Imamura} {et~al.}(1984){Imamura}, {Wolff}, \& {Durisen}}]{imam84}
{Imamura}, J.~N., {Wolff}, M.~T., \& {Durisen}, R.~H. 1984, \apj, 276, 667

\bibitem[{{Inayoshi} \& {Omukai}(2012)}]{io12}
{Inayoshi}, K. \& {Omukai}, K. 2012, \mnras, 422, 2539

\bibitem[{{Jeon} {et~al.}(2012){Jeon}, {Pawlik}, {Greif}, {Glover}, {Bromm},
  {Milosavljevi{\'c}}, \& {Klessen}}]{jeon11}
{Jeon}, M., {Pawlik}, A.~H., {Greif}, T.~H., {Glover}, S.~C.~O., {Bromm}, V.,
  {Milosavljevi{\'c}}, M., \& {Klessen}, R.~S. 2012, \apj, 754, 34

\bibitem[{{Joggerst} {et~al.}(2010){Joggerst}, {Almgren}, {Bell}, {Heger},
  {Whalen}, \& {Woosley}}]{jet09b}
{Joggerst}, C.~C., {Almgren}, A., {Bell}, J., {Heger}, A., {Whalen}, D., \&
  {Woosley}, S.~E. 2010, \apj, 709, 11

\bibitem[{{Joggerst} \& {Whalen}(2011)}]{jw11}
{Joggerst}, C.~C. \& {Whalen}, D.~J. 2011, \apj, 728, 129

\bibitem[{{Johnson} \& {Bromm}(2007)}]{jb07b}
{Johnson}, J.~L. \& {Bromm}, V. 2007, \mnras, 374, 1557

\bibitem[{{Johnson} {et~al.}(2013{\natexlab{a}}){Johnson}, {Dalla}, \&
  {Khochfar}}]{jdk12}
{Johnson}, J.~L., {Dalla}, V.~C., \& {Khochfar}, S. 2013{\natexlab{a}}, \mnras,
  428, 1857

\bibitem[{{Johnson} {et~al.}(2008){Johnson}, {Greif}, \& {Bromm}}]{jgb08}
{Johnson}, J.~L., {Greif}, T.~H., \& {Bromm}, V. 2008, \mnras, 388, 26

\bibitem[{{Johnson} {et~al.}(2009){Johnson}, {Greif}, {Bromm}, {Klessen}, \&
  {Ippolito}}]{jlj09}
{Johnson}, J.~L., {Greif}, T.~H., {Bromm}, V., {Klessen}, R.~S., \& {Ippolito},
  J. 2009, \mnras, 399, 37

\bibitem[{{Johnson} {et~al.}(2013{\natexlab{b}}){Johnson}, {Whalen}, {Even},
  {Fryer}, {Heger}, {Smidt}, \& {Chen}}]{jet13a}
{Johnson}, J.~L., {Whalen}, D.~J., {Even}, W., {Fryer}, C.~L., {Heger}, A.,
  {Smidt}, J., \& {Chen}, K.-J. 2013{\natexlab{b}}, arXiv:1304.4601

\bibitem[{{Johnson} {et~al.}(2012){Johnson}, {Whalen}, {Fryer}, \&
  {Li}}]{jlj12a}
{Johnson}, J.~L., {Whalen}, D.~J., {Fryer}, C.~L., \& {Li}, H. 2012, \apj, 750,
  66

\bibitem[{{Johnson} {et~al.}(2013{\natexlab{c}}){Johnson}, {Whalen}, {Li}, \&
  {Holz}}]{jet13}
{Johnson}, J.~L., {Whalen}, D.~J., {Li}, H., \& {Holz}, D.~E.
  2013{\natexlab{c}}, \apj, 771, 116

\bibitem[{{Kasen} {et~al.}(2011){Kasen}, {Woosley}, \& {Heger}}]{kasen11}
{Kasen}, D., {Woosley}, S.~E., \& {Heger}, A. 2011, \apj, 734, 102

\bibitem[{{Kitayama} \& {Yoshida}(2005)}]{ky05}
{Kitayama}, T. \& {Yoshida}, N. 2005, \apj, 630, 675

\bibitem[{{Kitayama} {et~al.}(2004){Kitayama}, {Yoshida}, {Susa}, \&
  {Umemura}}]{ket04}
{Kitayama}, T., {Yoshida}, N., {Susa}, H., \& {Umemura}, M. 2004, \apj, 613,
  631

\bibitem[{{Krti{\v c}ka} \& {Kub{\'a}t}(2006)}]{kk06}
{Krti{\v c}ka}, J. \& {Kub{\'a}t}, J. 2006, \aap, 446, 1039

\bibitem[{{Kudritzki}(2000)}]{Kudritzki00}
{Kudritzki}, R. 2000, in The First Stars, ed. {A.~Weiss, T.~G.~Abel, \&
  V.~Hill}, 127--+

\bibitem[{{Lai} {et~al.}(2008){Lai}, {Bolte}, {Johnson}, {Lucatello}, {Heger},
  \& {Woosley}}]{Lai2008}
{Lai}, D.~K., {Bolte}, M., {Johnson}, J.~A., {Lucatello}, S., {Heger}, A., \&
  {Woosley}, S.~E. 2008, \apj, 681, 1524

\bibitem[{{Latif} {et~al.}(2013{\natexlab{a}}){Latif}, {Schleicher}, {Schmidt},
  \& {Niemeyer}}]{latif13c}
{Latif}, M.~A., {Schleicher}, D.~R.~G., {Schmidt}, W., \& {Niemeyer}, J.
  2013{\natexlab{a}}, \mnras, 433, 1607

\bibitem[{{Latif} {et~al.}(2013{\natexlab{b}}){Latif}, {Schleicher}, {Schmidt},
  \& {Niemeyer}}]{latif13a}
---. 2013{\natexlab{b}}, \mnras, 430, 588

\bibitem[{{Li}(2011)}]{li11}
{Li}, Y. 2011, arXiv:1109.3442

\bibitem[{{Lippai} {et~al.}(2009){Lippai}, {Frei}, \& {Haiman}}]{lfh09}
{Lippai}, Z., {Frei}, Z., \& {Haiman}, Z. 2009, \apj, 701, 360

\bibitem[{{Mackey} {et~al.}(2003){Mackey}, {Bromm}, \& {Hernquist}}]{mbh03}
{Mackey}, J., {Bromm}, V., \& {Hernquist}, L. 2003, \apj, 586, 1

\bibitem[{{Madau}(1995)}]{madau95}
{Madau}, P. 1995, \apj, 441, 18

\bibitem[{{Magee} {et~al.}(1995){Magee}, {Abdallah}, {Clark}, {Cohen},
  {Collins}, {Csanak}, {Fontes}, {Gauger}, {Keady}, {Kilcrease}, \&
  {Merts}}]{oplib}
{Magee}, N.~H., {Abdallah}, Jr., J., {Clark}, R.~E.~H., {Cohen}, J.~S.,
  {Collins}, L.~A., {Csanak}, G., {Fontes}, C.~J., {Gauger}, A., {Keady},
  J.~J., {Kilcrease}, D.~P., \& {Merts}, A.~L. 1995, in Astronomical Society of
  the Pacific Conference Series, Vol.~78, Astrophysical Applications of
  Powerful New Databases, ed. {S.~J.~Adelman \& W.~L.~Wiese}, 51

\bibitem[{{Maio} {et~al.}(2011){Maio}, {Khochfar}, {Johnson}, \&
  {Ciardi}}]{maio11}
{Maio}, U., {Khochfar}, S., {Johnson}, J.~L., \& {Ciardi}, B. 2011, \mnras,
  414, 1145

\bibitem[{{McKee} \& {Tan}(2008)}]{tm08}
{McKee}, C.~F. \& {Tan}, J.~C. 2008, \apj, 681, 771

\bibitem[{{Meiksin} \& {Whalen}(2013)}]{mw12}
{Meiksin}, A. \& {Whalen}, D.~J. 2013, \mnras, 430, 2854

\bibitem[{{Mesler} {et~al.}(2012){Mesler}, {Whalen}, {Lloyd-Ronning}, {Fryer},
  \& {Pihlstr{\"o}m}}]{met12a}
{Mesler}, R.~A., {Whalen}, D.~J., {Lloyd-Ronning}, N.~M., {Fryer}, C.~L., \&
  {Pihlstr{\"o}m}, Y.~M. 2012, \apj, 757, 117

\bibitem[{{Milosavljevi{\'c}} {et~al.}(2009){Milosavljevi{\'c}}, {Bromm},
  {Couch}, \& {Oh}}]{milos09}
{Milosavljevi{\'c}}, M., {Bromm}, V., {Couch}, S.~M., \& {Oh}, S.~P. 2009,
  \apj, 698, 766

\bibitem[{{Montero} {et~al.}(2012){Montero}, {Janka}, \&
  {M{\"u}ller}}]{montero12}
{Montero}, P.~J., {Janka}, H.-T., \& {M{\"u}ller}, E. 2012, \apj, 749, 37

\bibitem[{{Moriya} {et~al.}(2010){Moriya}, {Yoshida}, {Tominaga}, {Blinnikov},
  {Maeda}, {Tanaka}, \& {Nomoto}}]{moriya10}
{Moriya}, T., {Yoshida}, N., {Tominaga}, N., {Blinnikov}, S.~I., {Maeda}, K.,
  {Tanaka}, M., \& {Nomoto}, K. 2010, in American Institute of Physics
  Conference Series, Vol. 1294, American Institute of Physics Conference
  Series, ed. {D.~J.~Whalen, V.~Bromm, \& N.~Yoshida}, 268--269

\bibitem[{{Moriya} {et~al.}(2013){Moriya}, {Blinnikov}, {Tominaga}, {Yoshida},
  {Tanaka}, {Maeda}, \& {Nomoto}}]{moriya12}
{Moriya}, T.~J., {Blinnikov}, S.~I., {Tominaga}, N., {Yoshida}, N., {Tanaka},
  M., {Maeda}, K., \& {Nomoto}, K. 2013, \mnras, 428, 1020

\bibitem[{{Mortlock} {et~al.}(2011){Mortlock}, {Warren}, {Venemans}, {Patel},
  {Hewett}, {McMahon}, {Simpson}, {Theuns}, {Gonz{\'a}les-Solares}, {Adamson},
  {Dye}, {Hambly}, {Hirst}, {Irwin}, {Kuiper}, {Lawrence}, \&
  {R{\"o}ttgering}}]{mort11}
{Mortlock}, D.~J., {Warren}, S.~J., {Venemans}, B.~P., {Patel}, M., {Hewett},
  P.~C., {McMahon}, R.~G., {Simpson}, C., {Theuns}, T., {Gonz{\'a}les-Solares},
  E.~A., {Adamson}, A., {Dye}, S., {Hambly}, N.~C., {Hirst}, P., {Irwin},
  M.~J., {Kuiper}, E., {Lawrence}, A., \& {R{\"o}ttgering}, H.~J.~A. 2011,
  \nat, 474, 616

\bibitem[{{Nagakura} {et~al.}(2012){Nagakura}, {Suwa}, \& {Ioka}}]{nsi12}
{Nagakura}, H., {Suwa}, Y., \& {Ioka}, K. 2012, \apj, 754, 85

\bibitem[{{Nakamura} \& {Umemura}(2001)}]{nu01}
{Nakamura}, F. \& {Umemura}, M. 2001, \apj, 548, 19

\bibitem[{{Ohkubo} {et~al.}(2009){Ohkubo}, {Nomoto}, {Umeda}, {Yoshida}, \&
  {Tsuruta}}]{ohk09}
{Ohkubo}, T., {Nomoto}, K., {Umeda}, H., {Yoshida}, N., \& {Tsuruta}, S. 2009,
  \apj, 706, 1184

\bibitem[{{Omukai} \& {Inutsuka}(2002)}]{oi02}
{Omukai}, K. \& {Inutsuka}, S.-i. 2002, \mnras, 332, 59

\bibitem[{{Omukai} \& {Palla}(2001)}]{op01}
{Omukai}, K. \& {Palla}, F. 2001, \apjl, 561, L55

\bibitem[{{Omukai} \& {Palla}(2003)}]{op03}
---. 2003, \apj, 589, 677

\bibitem[{{O'Shea} {et~al.}(2005){O'Shea}, {Abel}, {Whalen}, \&
  {Norman}}]{oet05}
{O'Shea}, B.~W., {Abel}, T., {Whalen}, D., \& {Norman}, M.~L. 2005, \apjl, 628,
  L5

\bibitem[{{O'Shea} \& {Norman}(2007)}]{on07}
{O'Shea}, B.~W. \& {Norman}, M.~L. 2007, \apj, 654, 66

\bibitem[{{O'Shea} \& {Norman}(2008)}]{on08}
---. 2008, \apj, 673, 14

\bibitem[{{Pan} {et~al.}(2012{\natexlab{a}}){Pan}, {Kasen}, \& {Loeb}}]{pan12a}
{Pan}, T., {Kasen}, D., \& {Loeb}, A. 2012{\natexlab{a}}, \mnras, 422, 2701

\bibitem[{{Pan} {et~al.}(2012{\natexlab{b}}){Pan}, {Loeb}, \& {Kasen}}]{pan12b}
{Pan}, T., {Loeb}, A., \& {Kasen}, D. 2012{\natexlab{b}}, \mnras, 423, 2203

\bibitem[{{Park} \& {Ricotti}(2011)}]{pm11}
{Park}, K. \& {Ricotti}, M. 2011, \apj, 739, 2

\bibitem[{{Park} \& {Ricotti}(2012)}]{pm12}
---. 2012, \apj, 747, 9

\bibitem[{{Park} \& {Ricotti}(2013)}]{pm13}
---. 2013, \apj, 767, 163

\bibitem[{{Pawlik} {et~al.}(2011){Pawlik}, {Milosavljevi{\'c}}, \&
  {Bromm}}]{pmb11}
{Pawlik}, A.~H., {Milosavljevi{\'c}}, M., \& {Bromm}, V. 2011, \apj, 731, 54

\bibitem[{{Pawlik} {et~al.}(2013){Pawlik}, {Milosavljevi{\'c}}, \&
  {Bromm}}]{pmb12}
---. 2013, \apj, 767, 59

\bibitem[{{Rakavy} \& {Shaviv}(1967)}]{rs67}
{Rakavy}, G. \& {Shaviv}, G. 1967, \apj, 148, 803

\bibitem[{{Regan} \& {Haehnelt}(2009)}]{rh09}
{Regan}, J.~A. \& {Haehnelt}, M.~G. 2009, \mnras, 396, 343

\bibitem[{{Reisswig} {et~al.}(2013){Reisswig}, {Ott}, {Abdikamalov}, {Haas},
  {Moesta}, \& {Schnetter}}]{reis13}
{Reisswig}, C., {Ott}, C.~D., {Abdikamalov}, E., {Haas}, R., {Moesta}, P., \&
  {Schnetter}, E. 2013, arXiv:1304.7787

\bibitem[{{Ritter} {et~al.}(2012){Ritter}, {Safranek-Shrader}, {Gnat},
  {Milosavljevi{\'c}}, \& {Bromm}}]{ritt12}
{Ritter}, J.~S., {Safranek-Shrader}, C., {Gnat}, O., {Milosavljevi{\'c}}, M.,
  \& {Bromm}, V. 2012, \apj, 761, 56

\bibitem[{{Scannapieco} {et~al.}(2005){Scannapieco}, {Madau}, {Woosley},
  {Heger}, \& {Ferrara}}]{sc05}
{Scannapieco}, E., {Madau}, P., {Woosley}, S., {Heger}, A., \& {Ferrara}, A.
  2005, \apj, 633, 1031

\bibitem[{{Sethi} {et~al.}(2010){Sethi}, {Haiman}, \& {Pandey}}]{sethi10}
{Sethi}, S., {Haiman}, Z., \& {Pandey}, K. 2010, \apj, 721, 615

\bibitem[{{Shang} {et~al.}(2010){Shang}, {Bryan}, \& {Haiman}}]{sbh10}
{Shang}, C., {Bryan}, G.~L., \& {Haiman}, Z. 2010, \mnras, 402, 1249

\bibitem[{{Shi} {et~al.}(1998){Shi}, {Fuller}, \& {Halzen}}]{sfh98}
{Shi}, X., {Fuller}, G.~M., \& {Halzen}, F. 1998, Physical Review Letters, 81,
  5722

\bibitem[{{Smith} \& {Sigurdsson}(2007)}]{ss07}
{Smith}, B.~D. \& {Sigurdsson}, S. 2007, \apjl, 661, L5

\bibitem[{{Smith} {et~al.}(2009){Smith}, {Turk}, {Sigurdsson}, {O'Shea}, \&
  {Norman}}]{bsmith09}
{Smith}, B.~D., {Turk}, M.~J., {Sigurdsson}, S., {O'Shea}, B.~W., \& {Norman},
  M.~L. 2009, \apj, 691, 441

\bibitem[{{Smith} {et~al.}(2011){Smith}, {Glover}, {Clark}, {Greif}, \&
  {Klessen}}]{sm11}
{Smith}, R.~J., {Glover}, S.~C.~O., {Clark}, P.~C., {Greif}, T., \& {Klessen},
  R.~S. 2011, \mnras, 414, 3633

\bibitem[{{Stacy} {et~al.}(2010){Stacy}, {Greif}, \& {Bromm}}]{stacy10}
{Stacy}, A., {Greif}, T.~H., \& {Bromm}, V. 2010, \mnras, 403, 45

\bibitem[{{Stacy} {et~al.}(2012){Stacy}, {Greif}, \& {Bromm}}]{stacy12}
---. 2012, \mnras, 422, 290

\bibitem[{{Su} {et~al.}(2011){Su}, {Stiavelli}, {Oesch}, {Trenti}, {Bergeron},
  {Bradley}, {Carollo}, {Dahlen}, {Ferguson}, {Giavalisco}, {Koekemoer},
  {Lilly}, {Lucas}, {Mobasher}, {Panagia}, \& {Pavlovsky}}]{su11}
{Su}, J., {Stiavelli}, M., {Oesch}, P., {Trenti}, M., {Bergeron}, E.,
  {Bradley}, L., {Carollo}, M., {Dahlen}, T., {Ferguson}, H.~C., {Giavalisco},
  M., {Koekemoer}, A., {Lilly}, S., {Lucas}, R.~A., {Mobasher}, B., {Panagia},
  N., \& {Pavlovsky}, C. 2011, \apj, 738, 123

\bibitem[{{Surman} {et~al.}(2006){Surman}, {McLaughlin}, \& {Hix}}]{smh06}
{Surman}, R., {McLaughlin}, G.~C., \& {Hix}, W.~R. 2006, \apj, 643, 1057

\bibitem[{{Surman} {et~al.}(2008){Surman}, {McLaughlin}, {Ruffert}, {Janka}, \&
  {Hix}}]{sm08}
{Surman}, R., {McLaughlin}, G.~C., {Ruffert}, M., {Janka}, H.-T., \& {Hix},
  W.~R. 2008, \apjl, 679, L117

\bibitem[{{Susa}(2013)}]{susa13}
{Susa}, H. 2013, \apj, 773, 185

\bibitem[{{Suwa} \& {Ioka}(2011)}]{suwa11}
{Suwa}, Y. \& {Ioka}, K. 2011, \apj, 726, 107

\bibitem[{{Tan} \& {McKee}(2004)}]{tm04}
{Tan}, J.~C. \& {McKee}, C.~F. 2004, \apj, 603, 383

\bibitem[{{Tanaka} {et~al.}(2012){Tanaka}, {Moriya}, {Yoshida}, \&
  {Nomoto}}]{tet12}
{Tanaka}, M., {Moriya}, T.~J., {Yoshida}, N., \& {Nomoto}, K. 2012, \mnras,
  422, 2675

\bibitem[{{Tanaka} \& {Haiman}(2009)}]{th09}
{Tanaka}, T. \& {Haiman}, Z. 2009, \apj, 696, 1798

\bibitem[{{Tominaga} {et~al.}(2011){Tominaga}, {Morokuma}, {Blinnikov},
  {Baklanov}, {Sorokina}, \& {Nomoto}}]{tomin11}
{Tominaga}, N., {Morokuma}, T., {Blinnikov}, S.~I., {Baklanov}, P., {Sorokina},
  E.~I., \& {Nomoto}, K. 2011, \apjs, 193, 20

\bibitem[{{Tornatore} {et~al.}(2007){Tornatore}, {Ferrara}, \&
  {Schneider}}]{tfs07}
{Tornatore}, L., {Ferrara}, A., \& {Schneider}, R. 2007, \mnras, 382, 945

\bibitem[{{Trenti} {et~al.}(2009){Trenti}, {Stiavelli}, \& {Michael
  Shull}}]{tss09}
{Trenti}, M., {Stiavelli}, M., \& {Michael Shull}, J. 2009, \apj, 700, 1672

\bibitem[{{Turk} {et~al.}(2009){Turk}, {Abel}, \& {O'Shea}}]{turk09}
{Turk}, M.~J., {Abel}, T., \& {O'Shea}, B. 2009, Science, 325, 601

\bibitem[{{Vasiliev} {et~al.}(2012){Vasiliev}, {Vorobyov}, {Matvienko},
  {Razoumov}, \& {Shchekinov}}]{vas12}
{Vasiliev}, E.~O., {Vorobyov}, E.~I., {Matvienko}, E.~E., {Razoumov}, A.~O., \&
  {Shchekinov}, Y.~A. 2012, Astronomy Reports, 56, 895

\bibitem[{{Vink} {et~al.}(2001){Vink}, {de Koter}, \& {Lamers}}]{Vink01}
{Vink}, J.~S., {de Koter}, A., \& {Lamers}, H.~J.~G.~L.~M. 2001, \aap, 369, 574

\bibitem[{{Volonteri} \& {Begelman}(2010)}]{vb10}
{Volonteri}, M. \& {Begelman}, M.~C. 2010, \mnras, 409, 1022

\bibitem[{{Weaver} {et~al.}(1978){Weaver}, {Zimmerman}, \&
  {Woosley}}]{Weaver1978}
{Weaver}, T.~A., {Zimmerman}, G.~B., \& {Woosley}, S.~E. 1978, \apj, 225, 1021

\bibitem[{{Weinmann} \& {Lilly}(2005)}]{wl05}
{Weinmann}, S.~M. \& {Lilly}, S.~J. 2005, \apj, 624, 526

\bibitem[{{Whalen} {et~al.}(2004){Whalen}, {Abel}, \& {Norman}}]{wan04}
{Whalen}, D., {Abel}, T., \& {Norman}, M.~L. 2004, \apj, 610, 14

\bibitem[{{Whalen} {et~al.}(2010){Whalen}, {Hueckstaedt}, \&
  {McConkie}}]{wet10}
{Whalen}, D., {Hueckstaedt}, R.~M., \& {McConkie}, T.~O. 2010, \apj, 712, 101

\bibitem[{{Whalen} \& {Norman}(2006)}]{wn06}
{Whalen}, D. \& {Norman}, M.~L. 2006, \apjs, 162, 281

\bibitem[{{Whalen} \& {Norman}(2008{\natexlab{a}})}]{wn08b}
---. 2008{\natexlab{a}}, \apj, 673, 664

\bibitem[{{Whalen} {et~al.}(2008{\natexlab{a}}){Whalen}, {O'Shea}, {Smidt}, \&
  {Norman}}]{wet08b}
{Whalen}, D., {O'Shea}, B.~W., {Smidt}, J., \& {Norman}, M.~L.
  2008{\natexlab{a}}, \apj, 679, 925

\bibitem[{{Whalen} {et~al.}(2008{\natexlab{b}}){Whalen}, {Prochaska}, {Heger},
  \& {Tumlinson}}]{wet08c}
{Whalen}, D., {Prochaska}, J.~X., {Heger}, A., \& {Tumlinson}, J.
  2008{\natexlab{b}}, \apj, 682, 1114

\bibitem[{{Whalen} {et~al.}(2008{\natexlab{c}}){Whalen}, {van Veelen},
  {O'Shea}, \& {Norman}}]{wet08a}
{Whalen}, D., {van Veelen}, B., {O'Shea}, B.~W., \& {Norman}, M.~L.
  2008{\natexlab{c}}, \apj, 682, 49

\bibitem[{{Whalen}(2012)}]{dw12}
{Whalen}, D.~J. 2012, arXiv:1209.4688

\bibitem[{{Whalen} {et~al.}(2012){Whalen}, {Even}, {Frey}, {Johnson},
  {Lovekin}, {Fryer}, {Stiavelli}, {Holz}, {Heger}, {Woosley}, \&
  {Hungerford}}]{wet12b}
{Whalen}, D.~J., {Even}, W., {Frey}, L.~H., {Johnson}, J.~L., {Lovekin}, C.~C.,
  {Fryer}, C.~L., {Stiavelli}, M., {Holz}, D.~E., {Heger}, A., {Woosley},
  S.~E., \& {Hungerford}, A.~L. 2012, arXiv:1211.4979

\bibitem[{{Whalen} {et~al.}(2013{\natexlab{a}}){Whalen}, {Even}, {Lovekin},
  {Fryer}, {Stiavelli}, {Roming}, {Cooke}, {Pritchard}, {Holz}, \&
  {Knight}}]{wet12e}
{Whalen}, D.~J., {Even}, W., {Lovekin}, C.~C., {Fryer}, C.~L., {Stiavelli}, M.,
  {Roming}, P.~W.~A., {Cooke}, J., {Pritchard}, T.~A., {Holz}, D.~E., \&
  {Knight}, C. 2013{\natexlab{a}}, \apj, 768, 195

\bibitem[{{Whalen} \& {Fryer}(2012)}]{wf12}
{Whalen}, D.~J. \& {Fryer}, C.~L. 2012, \apjl, 756, L19

\bibitem[{{Whalen} {et~al.}(2013{\natexlab{b}}){Whalen}, {Fryer}, {Holz},
  {Heger}, {Woosley}, {Stiavelli}, {Even}, \& {Frey}}]{wet12a}
{Whalen}, D.~J., {Fryer}, C.~L., {Holz}, D.~E., {Heger}, A., {Woosley}, S.~E.,
  {Stiavelli}, M., {Even}, W., \& {Frey}, L.~H. 2013{\natexlab{b}}, \apjl, 762,
  L6

\bibitem[{{Whalen} {et~al.}(2013{\natexlab{c}}){Whalen}, {Joggerst}, {Fryer},
  {Stiavelli}, {Heger}, \& {Holz}}]{wet12c}
{Whalen}, D.~J., {Joggerst}, C.~C., {Fryer}, C.~L., {Stiavelli}, M., {Heger},
  A., \& {Holz}, D.~E. 2013{\natexlab{c}}, \apj, 768, 95

\bibitem[{{Whalen} {et~al.}(2013{\natexlab{d}}){Whalen}, {Johnson}, {Smidt},
  {Meiksin}, {Heger}, {Even}, \& {Fryer}}]{wet13a}
{Whalen}, D.~J., {Johnson}, J.~J., {Smidt}, J., {Meiksin}, A., {Heger}, A.,
  {Even}, W., \& {Fryer}, C.~L. 2013{\natexlab{d}}, arXiv:1305.6966

\bibitem[{{Whalen} {et~al.}(2013{\natexlab{e}}){Whalen}, {Johnson}, {Smidt},
  {Heger}, {Even}, \& {Fryer}}]{wet13b}
{Whalen}, D.~J., {Johnson}, J.~L., {Smidt}, J., {Heger}, A., {Even}, W., \&
  {Fryer}, C.~L. 2013{\natexlab{e}}, arXiv:1308.3278

\bibitem[{{Whalen} \& {Norman}(2008{\natexlab{b}})}]{wn08a}
{Whalen}, D.~J. \& {Norman}, M.~L. 2008{\natexlab{b}}, \apj, 672, 287

\bibitem[{{Whalen} {et~al.}(2013{\natexlab{f}}){Whalen}, {Xu}, {Johnson}, {Li},
  \& {Fryer}}]{wet13}
{Whalen}, D.~J., {Xu}, H., {Johnson}, J.~L., {Li}, H., \& {Fryer}, C.~L.
  2013{\natexlab{f}}, \apj, in prep

\bibitem[{{Wise} \& {Abel}(2005)}]{wa05}
{Wise}, J.~H. \& {Abel}, T. 2005, \apj, 629, 615

\bibitem[{{Wise} \& {Abel}(2007)}]{wa07}
---. 2007, \apj, 671, 1559

\bibitem[{{Wise} \& {Abel}(2008)}]{wa08a}
---. 2008, \apj, 684, 1

\bibitem[{{Wise} {et~al.}(2008){Wise}, {Turk}, \& {Abel}}]{wta08}
{Wise}, J.~H., {Turk}, M.~J., \& {Abel}, T. 2008, \apj, 682, 745

\bibitem[{{Wise} {et~al.}(2012){Wise}, {Turk}, {Norman}, \& {Abel}}]{wise12}
{Wise}, J.~H., {Turk}, M.~J., {Norman}, M.~L., \& {Abel}, T. 2012, \apj, 745,
  50

\bibitem[{{Wolcott-Green} {et~al.}(2011){Wolcott-Green}, {Haiman}, \&
  {Bryan}}]{whb11}
{Wolcott-Green}, J., {Haiman}, Z., \& {Bryan}, G.~L. 2011, \mnras, 418, 838

\bibitem[{{Woosley} {et~al.}(2002){Woosley}, {Heger}, \&
  {Weaver}}]{Woosley2002}
{Woosley}, S.~E., {Heger}, A., \& {Weaver}, T.~A. 2002, Reviews of Modern
  Physics, 74, 1015

\bibitem[{{Yoshida} {et~al.}(2008){Yoshida}, {Omukai}, \& {Hernquist}}]{y08}
{Yoshida}, N., {Omukai}, K., \& {Hernquist}, L. 2008, Science, 321, 669

\end{thebibliography}

\end{document}